# Sound in occupied open-plan offices: Objective metrics with a review of historical perspectives

Manuj Yadav [a,b,*], Densil Cabrera [a], Jungsoo Kim [a], Janina Fels [b], Richard de Dear [a]

[a] Sydney School of Architecture, Design and Planning, The University of Sydney, Sydney, NSW 2006, Australia
[b] Institute of Technical Acoustics, RWTH Aachen University, Kopernikusstraße 5, 52074 Aachen, Germany

ABSTRACT

Open-plan offices (OPOs) have been around for more than half a century now, chronicling the vicissitudes of workplace topography amongst other factors. This paper addresses one such factor – the sound environment in *occupied* OPOs in relation to several objective workplace parameters, using measurements in contemporary OPOs and comparisons with studies over the last 50 years. Omnidirectional and binaural sound measurements were conducted in 43 offices during typical working hours. The results describe variation in several acoustic and psychoacoustic metrics, and present statistical models that predict these metrics as a function of the number of workstations in offices. $L_{A,eq}$ of 53.6 dB is typical for occupied OPOs, with spectral slope of approximately −4 dB/octave. $L_{A,eq}$ values do not vary much over the workplace parameters studied (e.g., floor plate area, work activity, etc), except for −2.7 dB and −4.1 dB differences between offices with/without carpeting, and offices with ceiling absorption but with/without carpeting, respectively; most likely from reduced floor impact noise leading to speech level reduction. Sound fluctuation, as characterised by the metric Noise Climate (*NCl*: $L_{A10}$ – $L_{A90}$) and the psychoacoustic Fluctuation Strength (*FS*), decreases significantly with increasing number of workstations in OPOs. This suggests lesser auditory distraction in larger offices, which needs further investigation. In terms of historical trends, OPOs have become quieter over the years, especially background noise quantified as $L_{A90}$, although there are several subtleties. Overall, current findings can inform several OPO design perspectives including policy documents, provide values for laboratory simulations of OPO acoustic environments, help interpret subjective impressions of OPO occupants, etc.

## 1. Introduction

Acoustic measurements in occupied open-plan offices (OPOs) span a period of 50 years starting from the 1970s, a period that marked an increase in the popularity of the open-plan design [1], to the current year 2020 [2,3]. Appendix A provides relevant details of previous studies reporting results from measurements in occupied OPOs over the last 50 years. Most studies report A-weighted equivalent energy sound pressure levels, $L_{A,eq}$, in decibels (dB) over a wide range of measurement durations and locations in offices. Many studies also report various A-weighted percentile levels ($L_{A5}$, $L_{A10}$, $L_{A50}$, $L_{A90}$, etc.) to represent the statistical time-dependence of sound, e.g., $L_{A90}$ represents the A-weighted sound pressure level exceeded 90% of the time. Some studies also include certain A-weighted levels combined in several metrics to encapsulate the notion of peaks/fluctuations in the sound relative to a baseline; the latter usually being $L_{A90}$ to represent the background noise in an occupied office. Such level-based fluctuation metrics are similar in principle, and include the *peak index* [4], *office noise index* [5,6], *noise pollution level* [5–7], *noise climate* [8], and $M_{A,eq}$ [2]. Some studies have reported reverberation times ($T$ in seconds) [2,5], and psychoacoustic metrics [9] (e.g., loudness, etc.) [6,10,11]. The other major group of related metrics include those derived from octave/one-third octave band spectra such as *noise rating*, *preferred noise criterion*, balance between spectral regions, etc [2,6,10–14].

These previous studies were conducted in a total of 9 countries, predominantly European, with some in North America, and Asia; within a variety of office types, e.g., landscaped, cellular, etc., and work themes (monothematic, mixed-function, etc.); a wide range of floor plate areas (200–3809 $m^2$; not all studies reported office sizes); and using omnidirectional microphones (when reported) where the microphone placement strategy varied between studies. Over the 50-year period that these studies span, substantial

* Corresponding author at: Institute of Technical Acoustics, RWTH Aachen University, Room 104, Kopernikusstraße 5, 52074 Aachen, Germany.

E-mail address: manuj.yadav@sydney.edu.au (M. Yadav).

changes can safely be assumed for several aspects of the OPO sound environment. These changes include quieter heating, ventilation and air-conditioning (HVAC) systems, printers and office machinery in general, enhanced glazing systems and reduced external noise, computers replacing typewriters, different telephone rings, etc. Besides, work cultures within and between countries, companies, and over time are likely to have undergone changes. The confluence of these myriad factors as well as a paucity of crucial details in previous studies (e.g., measurement details, office sizes, noise sources, etc.) hampers attempts to summarise the OPO acoustic literature without excessive caveats, which poses limitations for advancing the science of OPO acoustics.

Hence, one of the main aims of this paper is to provide a comprehensive quantitative assessment of the physical sound environment in a representative sample of contemporary OPOs during working hours, using a consistent method, and for several key workplace factors (workstation numbers/density, office types, etc.). An associated goal involves determining whether a relationship can be established between the physical sound environment and the workplace factors. These results will be useful in the design and comparison of OPO sound environments in actual offices and in laboratory studies, informing policy documents, acoustic consulting, etc.

Another aim of this paper is to summarise the findings of previous occupied OPO studies, to chart the changes in the OPO acoustic environment over its history. The motivation here is to not only contextualise findings from the current data, but to also provide relevant details for a diverse audience within the broader scope of multidisciplinary OPO research. This addresses researchers within the disciplines of cognitive psychology, property management, ergonomics, etc., who may find the summary presented here useful within their respective fields, or within interdisciplinary studies. Finally, measurement results using both omnidirectional microphones and binaural dummy head are presented. While omnidirectional measurements are common in OPO studies to capture the 'ambient' level, binaural measurements capture some of the effect of the human head including the outer ears, hence the aim here is to provide results that are arguably more representative of hearing conditions in offices.

To address these aims for a wide range of OPOs, the current paper has a fairly inclusive operating definition – it considers OPOs as fully air-conditioned, medium-to large-sized office floors with workstations that are not separated by walls. Although the underlying philosophy of the open-plan design and implementation has undergone several iterations over the years [1], this broad scope allows the consideration of a whole gamut of factors such as territoriality (e.g., fixed workstations for employees, 'hot-desking' [15]), degree of openness (e.g., workstations within cubicles, or with limited to no partitions), total number and density of workstations, predominant activities (e.g., clerical, design work), etc., and combinations thereof, and the impact of these factors on acoustic and psychoacoustic metrics. Offices within industrial settings with a high proportion of machinery-based activities, laboratory studies, and smaller offices with less than four occupants are not considered here.

## 2. Materials and methods

### 2.1. Offices

Table 1 summarises key workplace environmental parameters of the 43 offices included in this study. These offices are within 9 buildings in metropolitan cities of Australia, labelled A-H. Offices are referred to as <building> . <office>, e.g., A.1, or simply by the building letter/office number. Offices in the same building are either on different floors or are non-contiguous units on the same floor but sufficiently different workplace and/or room acoustic environment to warrant separate analysis. All offices had centralised HVAC systems and none had sound masking systems.

Table 1 shows the number of workstations/employees and the workstation/employee density (number of workstations per 100 $m^2$) in the sample of offices. The percentage of employees with workstations allocated on a regular basis ranged from 20 to 100 % (*median*: 86%) if offices in two buildings are excluded. These two buildings (E and G; see supplementary material 1) include offices where none of the employees have a pre-allocated workstation. Such offices are labelled activity-based workplaces (ABW) or related terms [15], where employees can, in principle, choose a workspace that suits their particular activity. This may include, but is not limited to, choosing a workstation in the open-plan area, meeting rooms/collaboration areas (both open-plan and enclosed rooms), or choosing designated areas/rooms for concentrated working alone. Both the ABW buildings in the current sample (E, G) had several such areas that allowed working away from workstations, although it was still common to hold conversations at and between workstations within the open-plan areas.

Apart from the occasional complicated ceiling configuration (see supplementary material 1), uniform horizontal ceilings around 2.7 m in height were the most prevalent. Most ceilings had at least some sound absorptive treatment, and 30 offices had the entire ceiling covered with absorptive tiles. Most offices (32 in total) had little to no partitions/screens between workstations, except for computer screens. Four different types of partition were observed in use, labelled types N, I, II, II, and III, corresponding to no partition, or partitions enclosing the workstations from one, two, or three sides, respectively. Partition heights ranged from 1.1 to 1.6 m; further details are provided in supplementary material 1.

Almost all the offices had employees grouped in several departments, and Table 1 lists the primary workplace activities of offices (abbreviations in Table 1 used later in figures). Two out of the nine buildings (C and F) had non-academic university offices, while the rest were commercial offices. The surface area in Table 1 represents the portion of the entire floor plate consisting of OPOs only and does not include areas for building services (elevator lobbies, plant rooms), kitchens, enclosed rooms for meetings, personnel, etc., as long as these were not within the office perimeter. This may partly account for smaller office areas reported in Table 1 than some previous studies, e.g., [5,14,16]; also see Appendix A.

Besides background noise from ventilation and speech, the constituents of sound environments within office workplaces are understandably diverse. The following lists typical sounds that were noticed during measurement campaigns, in no particular order, and by no means exhaustive. The two broad groups of sounds in offices include task-specific non-speech sounds (mouse clicks, keyboard operation, shuffling items on work surfaces, writing on paper, papers furling, creaking from chairs and other furniture, computer – and printer–related noise), and occupants' movements and miscellaneous actions (moving chairs, footsteps, objects being dropped, coughing, sneezing, clearing throats, eating including sound from cutlery use, whistling, plastic wrappers, lift lobby sounds, doors being operated, and mobile/desk phone noise).

### 2.2. Measurements

The overall aim here was to conduct both binaural and omnidirectional measurements in offices *during typical working hours*, to be used for offline analysis; protocol approved by The University of Sydney Human Research Ethics Committee (Project: 2017/285), and signed by each building manager. The measurements were conducted during typical working hours of 09:00–

**Table 1**
Summary of several workplace parameters.

| **Summary statistics of 43 offices in 9 buildings (A-H)** | | | | | |
|---|---|---|---|---|---|
| **Parameter** | **Mean** | **Standard Deviation** | **Median** | **Median absolute deviation** | **Range** |
| Number of workstations | 35.6 | 17.5 | 30 | 14 | 12 – 78 |
| Workstation Density (per 100 m$^2$) | 12.1 | 5.6 | 11 | 6 | 4 – 24 |
| Ceiling height (m) | 3.3 | 1.4 | 2.7 | 0 | 2.2 – 7.6 |
| Surface area (m$^2$) | 201.2 | 140.2 | 174.4 | 93.3 | 53 – 719 |
| **Summary of categories (Total Number and Percentage)** | | | | | |
| Ceiling type | Absorptive = 30 (69.8%), Hard = 13 (30.2%) | | | | |
| Carpet | Yes = 34 (79.1%), No = 9 (20.9%) | | | | |
| ABW (Activity-based workplace) | Yes = 19 (44.2%), No = 24 (55.8%) | | | | |
| Partition type | Yes = 11 (25.6%), No = 32 (74.4%) | | | | |
| Surface area | (*value* $\leq$ 100) = 9 (20.9%), (101 $\leq$ *value* $\leq$ 200) = 17 (39.5%), (201 $\leq$ *value* $\leq$ 300) = 10 (23.3%), (*value* $\geq$ 301) = 7 (16.3%) | | | | |
| Work activities | Architecture, Design (Arch.) = 7 (16.3%), Policy (Plcy.) = 11 (25.6%), Engineering (Eng.) = 5 (11.6%), Management (Mgm.) = 18 (41.9%), Customer Service (CS) = 2 (4.7%) | | | | |

17:00. Since the schedules and duration for employees' lunch breaks showed a wide variation, measurements were conducted throughout the working day when appropriate.

For the binaural measurements, the Neumann KU100 (Berlin, Germany) dummy head was used, which models a human listener [17]. The signal chain included a laptop computer and an RME Babyface Pro interface (Haimhausen, Germany) for recording 2-channel audio at a sampling rate of 44.1 kHz. For the omnidirectional measurements, two types of sound level meters (SLM) were used: Brüel and Kjær (B&K) Type 2250 SLM with Type 4189 omnidirectional microphone (Nærum, Denmark) for buildings A-H, and both B&K and NTi XL2 SLM with M426 omnidirectional microphone (Schaan, Lichtenstein) for buildings H and I. In all offices, the dummy head was placed at various workstations at a height of 1.2 m (floor to ear canal entrance) and 0.5 m from the desk to represent a typical seating position, and at least 2 m from the walls. The SLMs were placed at the same workstations as the dummy head, but also at other workstations.

The overall measurement approach included sampling as many workstations as possible and deemed necessary for an adequate representation of the entire office, based on visual and aural inspection. This purposive sampling strategy was adapted to each office's circumstances to accommodate several factors: the measurement duration permitted/possible per office and per workstation, the size and homogeneity of partition and ceiling types, work activities etc. This meant shorter duration measurements at many workstations when there was a larger variation between locations within the office, rather than longer measurements at fewer locations, and represents an employee spending time at various locations in the office.

More specifically, for buildings A-E and I, the measurements were conducted over one day during which all the offices per building were measured using the dummy head. In these offices, measurements per workstation ranged from 15 min to 2 h to collect representative samples. For building F, dummy head measurements were conducted per office per day (2 h per workstation) and B&K SLMs were used to measure offices F.24-F.30 for a whole week. These measurements (i.e., offices F.24-F.30) represent the longest sampling duration in this study. For building G, dummy head measurements were conducted over two days, one day each for offices G.31-G.34 and G.35-G.38, and four B&K SLMs were used per office per day. For building H, dummy head measurements were conducted over one day for all the offices, and four B&K and four NTi SLMs were used at several workstations to sample each office for the entire day. *In total, all 43 offices were measured using a dummy head at various workstations, and 18 out of these 43 offices were also measured over at least an 8-hour period using SLMs at various workstations.*

Besides the measurements during working hours, several other measurements were performed in these offices. It included room acoustic measurements (according to ISO 3382-2 [18] and ISO 3382-3 [19]) in most of the offices outside of working hours, and detailed results for buildings A-H are presented in a previous paper [20]. Employee survey was also conducted, the results of which will be covered in a subsequent publication.

### 2.3. Data processing

For the binaural measurements, the values reported are based on a 4-hour averaging period, unless mentioned otherwise. This duration represented the most common measurement duration for all offices, and all statistical analyses use these 4-hour averaged values. However, results for other averaging durations, e.g., 15, 30 min, etc. are also reported in certain cases. Some offices were measured for longer than 4 h, in which case a continuous 4-hour sample was selected. Some offices were measured for slightly <4 h, in which case the remaining period was completed by appending 1 s samples selected at random with replacement from the measurement.

Overall, three categories of metrics were calculated:

i. Those *based on A-weighted sound pressure levels* (in decibels). These include $L_{A,eq,4h}$, several percentile levels ($L_{A10,4h}$, $L_{A50,4h}$, $L_{A90,4h}$), and level-based fluctuation metrics such as the peak index (*PI*): cumulative sum of levels exceeding the $L_{A,eq,4h}$ by 5, 10 and 15 dB [4]; office noise index (*ONI*): $L_{A90,\ 4h} + 2.4\times(L_{A10,4h} - L_{A90,4h}) - 14$ [5,6]; noise pollution level (*NPL*): $L_{A,eq,4h} + (L_{A10,4h} - L_{A90,4h})$) [5–7]; noise climate (*NCl*): $L_{A10,4h} - L_{A90,4h}$ [8], and $M_{A,eq}$: ($L_{A,eq,4h} - L_{A90,4h}$) [2].
ii. Those *based on octave/one-third octave band filtering*. Besides giving an averaged spectral representation, the filtered data can be used to derive certain metrics used in previous studies [6,10,14]. These metrics include the noise rating (*NR*) [21], room criterion Mark II (*RC*) [22], balanced noise-criterion (*NCB*) [23], and the difference between the A-weighted SPL averages of the *low* (16–63 Hz) and *high* (1000–4000 Hz) one-third octave band decibel levels (*Lo-Hi*) [24]. Since these noise metrics were developed for, and are used primarily for studying HVAC noise and/or noise during occupation without speech [25], the results from such metrics are not presented in the text, but are presented within supplementary material 2.
iii. Those *based on psychoacoustics*. These metrics include binaural loudness (*N* in sones [9]) calculated using Moore and Glasberg's time-varying binaural loudness model [26] with the middle ear transfer function presented in [27]; sharpness (*S* in acum [9]) calculated using [28]; roughness ($R$ in asper [9]) calculated using [29], fluctuation strength ($FS$ in vacil [9]) calculated using [30], and loudness fluctuation ($N_{Fluctuation}$, which is a measure of fluctuation of loudness values over time, calculated using [28]. For all metrics besides $N$, the value reported is the average value of the two ears. MATLAB® code for these psychoacoustic models is available from [31].

For the omnidirectional measurements, results for only category (i) and some for category (ii) are presented, along with reverberation times ($T_{30}$ in seconds; average of the 500 Hz and 1000 Hz octave bands).

### 2.4. Statistical analyses

Statistical analyses were done within the software R [32], using the packages *tidyverse* [33] for data management, testing distributional assumptions and generating graphics; using package *nlme* [34] for assessing the need for generalised mixed-effects models (GLMM) over generalised least-square models (GLM), and for assessing the significance of a fixed-effect compared to the null (intercept-only) model; and using package *robustlmm* [35] for fitting robust linear mixed-effects models (RLMM), wherein the robustness means minimising the influence of outliers.

The general form of the RLMM is shown in Eq. 1 (interaction models had an extra fixed-effect), where $y$ represents the dependent variable: metrics from types (i) and (iii) described in this section; $a$ represents the fixed-intercept, $b$ the fixed-effect slope; the $\varepsilon$ values represent the random-effects: varying slopes for the buildings, and the residual error.

$$y = a + bx + \varepsilon_{building} + \varepsilon_{residual} \quad (1)$$

For the *fixed-effect* ($x$) in Eq. 1, i.e., the predictor in the RLMM, all the key physical parameters describing the offices (Table 1), such as the number of workstations, workstation density, floor plate area, etc., were tested and reported when necessary. However, the number of workstations ($WS_n$) was chosen as the default fixed-effect for models reported here. This is because $WS_n$ is arguably the most salient and straightforward parameter to interpret out of the ones relevant in offices and is moreover highly correlated with other parameters ($R^2$ > 0.7). Besides the fixed-effect, the unsystematic *random-effect* due to autocorrelated and non-independent data – some offices were within the same building – was explicitly modelled in the RLMM by allowing the intercepts per *building* to vary independently.

Statistical significance of the fixed-effect was determined by comparing the log-likelihood values of the null (intercept-only) model with a model that included the fixed-effect. This comparison was done using a chi-square test ($\chi^2_{(\textbf{null vs. } x)}$), with $p$ < .05 chosen as the criterion for establishing statistical significance for the fixed-effect. Residuals of the models met parametric assumptions, i.e., linearity, normality, and homoscedasticity. The statistical significance of the fixed-effect slope ($b$ in Eq. 1) was determined by the respective 95% confidence interval (CI; calculated using Wald's test) not crossing the null value of 0. The goodness-of-fit of the RLMM is presented as the *conditional* $R^2$ value as defined in Eq. 2, wherein $\sigma_f^2$ is the variance explained by the fixed-effects, $\sigma_\alpha^2$ is the random-effects variance (normally distributed with mean zero), and $\sigma_\varepsilon^2$ is the residual-error variance. Hence, $R^2_{RLMM}$ describes the proportion of variance explained by both the fixed-and random-effects, which is more appropriate for mixed-effects models than the traditional $R^2$ used for GLMs [36].

$$R^2_{RLMM} = \frac{\sigma^2 + \sigma_\alpha^2}{\sigma^2 + \sigma_\alpha^2 + \sigma_\varepsilon^2} \quad (2)$$

## 3. Results

The following sections present results for each category of metrics that were calculated (Section 2.3). Table 2 presents summary of the various metrics considered in this paper. Values for all the metrics and workplace parameters are provided within supplementary material 3.

### 3.1. Binaural measurements

#### 3.1.1. Based on A-weighted sound pressure levels

The left panel of Fig. 1 shows summary of $L_{A,eq,4h}$, $L_{A10,4h}$, $L_{A50,4h}$, and $L_{A10,4h}$ values over the offices, which had ranges of 10.5 dB, 10.9 dB, 5.4 dB and 11.6 dB respectively. The right panel shows the $L_{A,eq,4h}$ values measured per second in the offices in buildings F-H in the form of respective empirical cumulative distribution (ECD) curves. These curves show the variety in the distribution of $L_{A,eq,4h}$ values for offices: while the curves are relatively similar for offices in Building H, offices of Building F and G show a much wider variation, both in shape and range of values. Buildings F and G are representative of the rest of the buildings, in that the ECD curves varied between offices of these buildings. In other words, most offices in the current sample of 43 offices, even those within the same building, showed variation from each other in terms of distribution of occupied levels (and other metrics). Hence, the current sample represents a rather wide range of offices, which may not otherwise be obvious as these 43 offices come from 9 buildings.

Fig. 2 shows the values per office in more detail for several A-weighted metrics. For these metrics, the effect of allowing the intercept per building to vary (random-effect) was significant, based on the metric-wise comparison of log-likelihood values of models without (i.e., GLM) and with (GLMM) this random-effect. As seen in Table 3, compared to the null (intercept-only) model, $WS_n$ was not a significant predictor of $L_{A,eq,4h}$ and $L_{A10,4h}$, office noise index (*ONI*), and noise pollution level *(NPL)*, but was a significant predictor of $L_{A50,4h}$ and $L_{A90,4h}$, noise climate (*NCl*), peak index (*PI*) and $M_{A,eq}$. The strongest RLMM is seen in Eq. 3 and the predicted values plotted in Fig. 2, wherein the *NCl* values (in dB(A)) are predicted with the number of workstations ($WS_n$) as the fixed-effect ($R^2_{RLMM}$ = 0.52).

$$NCl = 25.48 - 0.02 \times WS_n + 0.67_{building} + 0.83_{residual} \quad (3)$$

Similar models can be derived using $M_{A,eq,}$ which was the other strong predictor with a slightly lower $R^2_{RLMM}$ value than *NCl* (Table 2), and had a similar distribution of values as *NCl*, as seen in Fig. 2.

Fig. 3 provides another perspective of the $L_{A,eq,4h}$ and *NCl* values. Here, each row presents scatterplots for $L_{A,eq,4h}$ (left column) and *NCl* (right column) values with respect to the number of workstations, where offices are grouped according to several key workplace parameters, one parameter per row. Groupings based on other parameters such as ceiling heights, etc., can be derived using supplementary material 3.

Fig. 3 does not present large mean/median differences in the $L_{A,eq,4h}$ values between offices (see respective boxplots) with and without absorptive ceilings. There were three main groups of ceiling heights in the data: 2.7 m (the most typical), 2.8–3.5 m, and greater than 3.6 m for the complicated ceiling types. However, for the fixed-effect of ceiling absorption predicting $L_{A,eq,4h}$ values, the random-effect of these ceiling height groups, modelled as independently varying intercepts in a GLMM, was not significant, compared to a fixed-intercept GLM ($\chi^2$ (1) < $10^{-2}$, $p$ = .99). The inclusion of independently varying intercepts for different buildings was also not statistically significant in a GLMM vs. fixed-intercept GLM model ($\chi^2$ (1) = 0.41, $p$ = .52). Based on the GLM, the difference in $L_{A,eq,4h}$ in offices without and with ceiling absorption was 0.27 dB, which was non-significant with 95% CI: [-1.71, 2.25].

**Table 2**
Summary statistics of the metrics. See Section 2.3 for short descriptions of the metrics.

| Metric | Unit | Mean | SD | Median | MAD | Range |
|---|---|---|---|---|---|---|
| $L_{A,eq}$ | decibel (dB) | 53.56 | 3.01 | 53.67 | 3.27 | 48.28 – 58.83 |
| $L_{A10}$ | decibel (dB) | 57.00 | 3.15 | 57.03 | 3.95 | 51.58 – 62.46 |
| $L_{A50}$ | decibel (dB) | 47.01 | 3.02 | 46.70 | 3.62 | 41.89 – 53.29 |
| $L_{A90}$ | decibel (dB) | 32.07 | 3.00 | 31.87 | 3.56 | 27.11 – 38.67 |
| *ONI* | decibel (dB) | 91.90 | 4.23 | 91.36 | 4.12 | 85.14 – 102.21 |
| *NPL* | decibel (dB) | 78.49 | 3.78 | 78.09 | 3.34 | 72.2 – 87.3 |
| *PI* | | 11.04 | 2.83 | 10.90 | 2.87 | 3.47 – 17.36 |
| *NCl* | decibel (dB) | 24.93 | 1.42 | 24.43 | 0.91 | 22.61 – 30.2 |
| $M_{A,eq}$ | decibel (dB) | 21.49 | 1.59 | 21.09 | 1.20 | 18.38 – 27.61 |
| $N_{mean}$ | sone | 6.44 | 1.19 | 6.33 | 1.45 | 4.59 – 8.99 |
| $N_{max}$ | sone | 31.62 | 9.48 | 28.87 | 7.99 | 16.67 – 62.42 |
| $N_5$ | sone | 9.73 | 1.86 | 9.49 | 1.91 | 6.44 – 14.63 |
| $N_{90}$ | sone | 4.79 | 1.02 | 4.74 | 1.11 | 3.17 – 7.02 |
| $N_{Fluctuation}$ | | 2.88 | 0.33 | 2.87 | 0.25 | 1.7 – 3.51 |
| $S_{mean}$ | acum | 1.17 | 0.09 | 1.16 | 0.07 | 1.03 – 1.36 |
| $S_{max}$ | acum | 2.76 | 0.31 | 2.71 | 0.30 | 1.99 – 3.44 |
| $S_5$ | acum | 1.54 | 0.11 | 1.53 | 0.10 | 1.24 – 1.75 |
| $S_{90}$ | acum | 3.98 | 2.17 | 4.67 | 1.68 | 0.88 – 7.39 |
| *FS* | vacil | 0.39 | 0.11 | 0.36 | 0.10 | 0.07 – 0.61 |
| $R_{mean}$ | asper | 0.08 | 0.02 | 0.08 | 0.02 | 0.05 – 0.16 |
| $R_{max}$ | asper | 4.47 | 1.49 | 4.61 | 1.68 | 1.24 – 7.13 |
| $R_5$ | asper | 0.02 | 0.00 | 0.02 | 0.01 | 0.01 – 0.02 |
| $R_{90}$ | asper | 0.14 | 0.07 | 0.13 | 0.02 | 0.08 – 0.38 |
| $T_{30}$ | seconds | 0.53 | 0.16 | 0.50 | 0.15 | 0.3 – 1.2 |

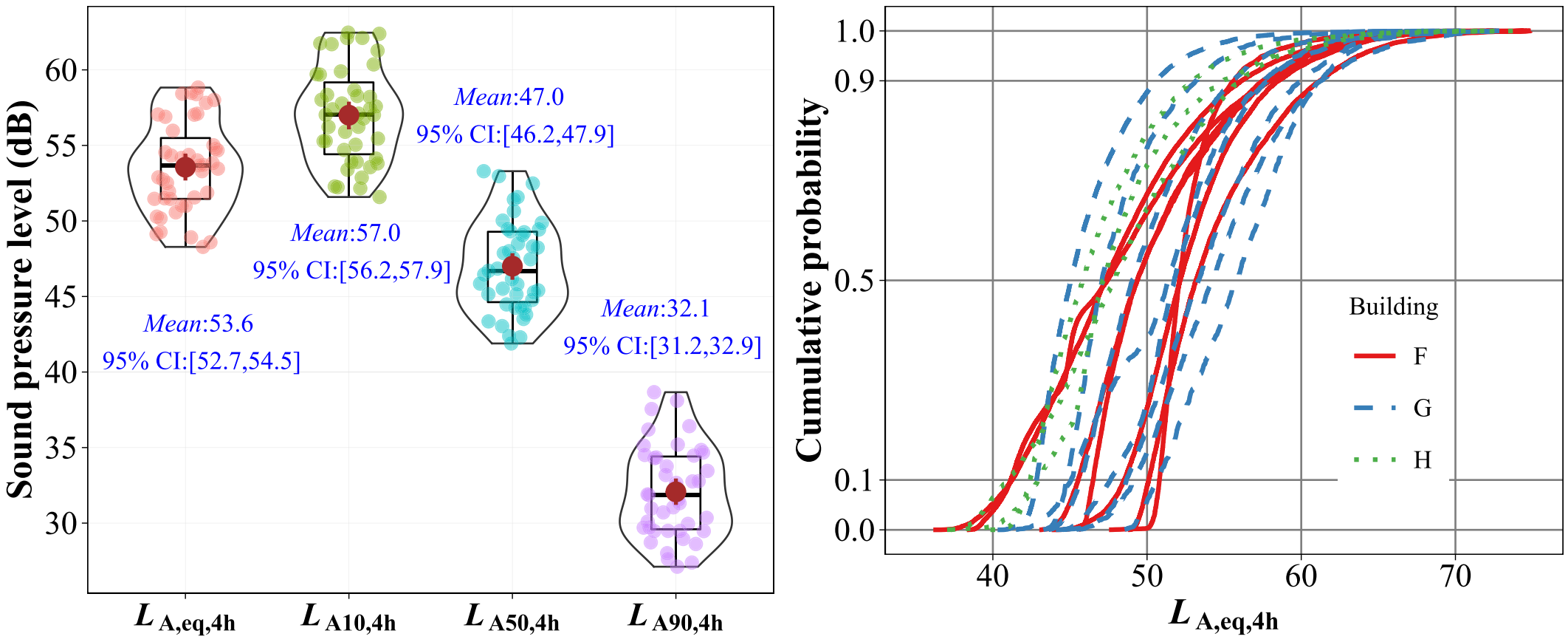

**Fig. 1.** (Left panel) Box and violin plots for the A-weighted metrics, along with individual values jittered along the *x*-axis, and summary statistics for all offices. (Right Panel) Examples of empirical cumulative probability distribution functions for the offices in three buildings (F-H).

Since none of the offices in the sample had appreciable wall absorption, the main absorptive surfaces in the offices were floors and ceilings. The interaction of the ceiling absorption and carpet groups was significant ($\chi^2$ (1) = 5.88, $p < .05$) based on comparison of GLMs with and without the interaction effect. The random-effect due to independently varying intercepts and slopes for the three ceiling height groups was not significant ($\chi^2$ (1) = 1.09, $p$ = .30, and $\chi^2$ (2) = 0.39, $p$ = .82, respectively), compared to the GLM without these random-effects. The results for the significant GLM with ceiling absorption and carpet groups as predictors, along with their interactions, are presented in Fig. 4. The absence of carpet did not lead to a statistically significant change in predicted $L_{A,eq,4h}$ for offices with no ceiling absorption (-1.23 dB; 95% CI: [-4.51, 2.05 dB]). However, for offices with ceiling absorption, the predicted $L_{A,eq,4h}$ without carpeting was significantly higher (Fig. 4) than for those with carpeting by 4.10 dB (95% CI: [1.01, 7.19]).

For the remaining parameters in Fig. 3, based on RLMMs with independently varying intercepts for the different buildings, the $L_{A,eq,4h}$ difference in offices without and with carpets was statistically significant and was −2.67 dB (95% CI: [-5.00, −0.35]). None of the other parameters had statistically significant $L_{A,eq,4h}$ differences between their respective groups. The $L_{A,eq,4h}$ difference between offices without and with partitions was 1.05 dB (95% CI: [-1.57, 3.67). For the prominent workplace activity, the $L_{A,eq,4h}$ differences were compared using orthogonal contrasts. The $L_{A,eq,4h}$ difference in Customer Service (CS.) offices with the rest of the offices was 0.07 dB (95% CI: [-0.92, 1.06]), between offices of Architecture and Design (Arch.) firms and the remaining offices (excluding CS offices) was −0.57 dB (95% CI: [-1.29, 0.16]), between Policy and Administration offices and the remaining offices (excluding CS and Arch. offices) was 0.18 dB (95% CI: [-0.71, 1.08]), and between offices of Engineering firms and Management firms was −0.02 dB (95% CI: [-1.74, 1.70]). The $L_{A,eq,4h}$ difference between offices that were not and were activity-based workplaces was 0.62 dB (95% CI: [-1.96, 3.20]). For the floor plate area categories, the $L_{A,eq,4h}$ differences were compared using orthogonal contrasts. The $L_{A,eq,4h}$ difference between offices with area<100 m$^2$ and the other offices was 0.42 dB (95% CI: [-0.04, 0.88]), between offices with areas between 101–200 m$^2$ and the remaining offices was 0.38 dB (95% CI: [-0.31, 1.07]), between offices with areas between 201–300 m$^2$ and areas greater than 301 m$^2$ was −0.45 dB (95% CI: [-1.90, 1.00]).

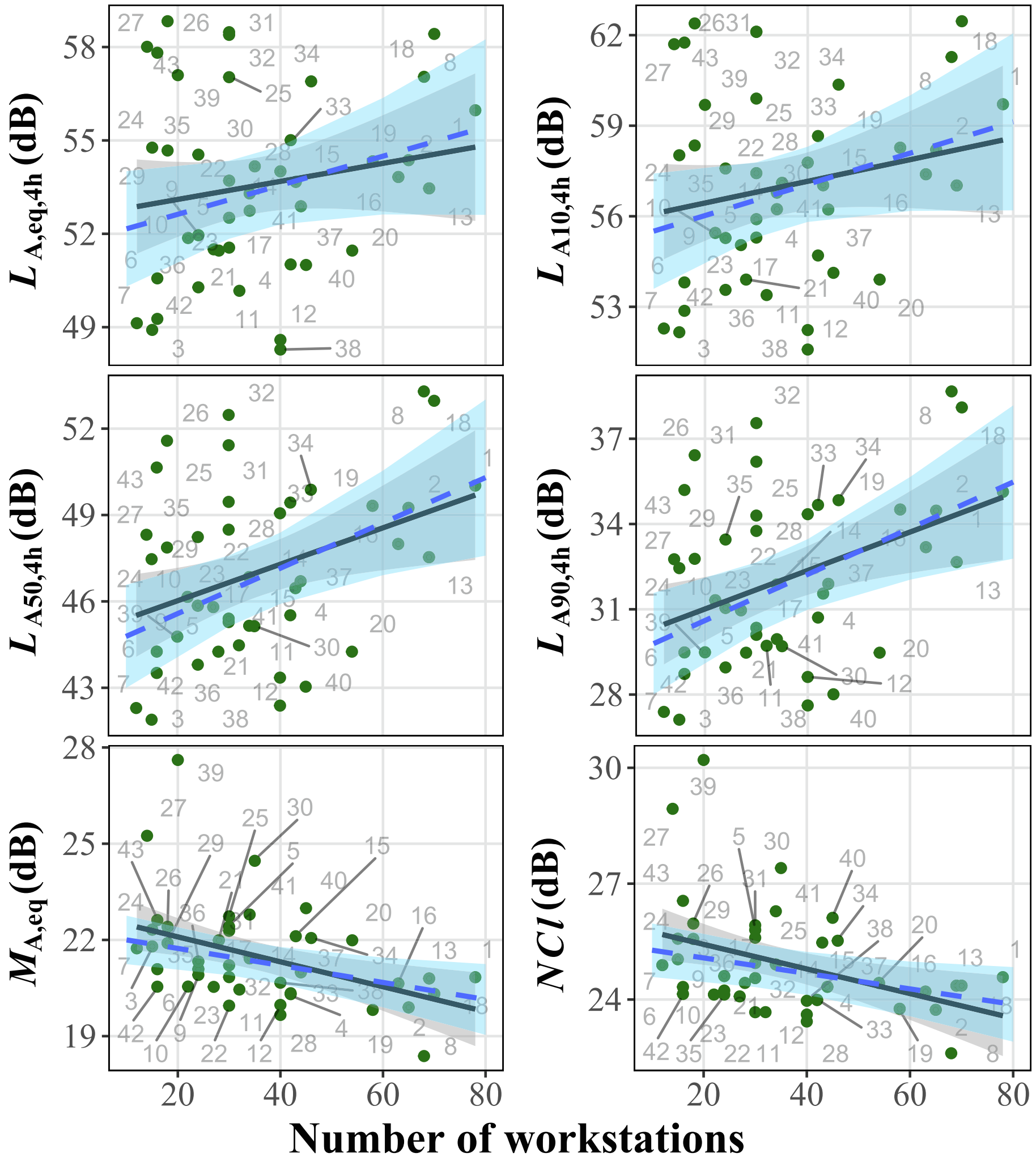

**Fig. 2.** Scatterplots for A-weighted metrics for the binaural measurements (left and right ears power averaged), where each data point is labelled with its office number. Per metric, two regression lines are presented: GLM fit as a solid line, and a dotted line connecting the predicted values using the RLMM fit (Table 2); both lines are presented with their respective shaded 95% confidence interval. Noise climate (*NCl*): $L_{A10,4h}$ - $L_{A90,4h}$ [8], and $M_{A,eq}$: ($L_{A,eq,4h}$ - $L_{A90,4h}$) [2].

**Table 3**
RLMM parameters of the form shown in Eq. 1 for various metrics based on A-weighted SPL values. The fixed-effect (***x***) in these models was the number of workstations ($WS_n$). Column 1 lists the dependent variable (***y***). Column 2–3 list the fixed-intercept (***a***) and fixed-effect slope (***b***) values, respectively. Column 4 lists the goodness-of-fit comparison of the null model vs. model with the fixed-effect ($\chi^2_{\text{null vs. } x}$ with 1 degree of freedom) and the respective ***p*-value**. Rows with significant $\chi^2_{\text{null vs. } x}$ values ($p < .05$) highlighted in bold. Columns 5–6 list the random-effects. Column 7 lists the goodness-of-fit of the model ($R^2_{\text{RLMM}}$), as defined in Eq. 2.

| *y* | *a* [95% CI] | *b* [95% CI] | $\chi^2_{\text{null vs. x}}(1)$, *p-value* | $\varepsilon_{\text{building}}$ | $\varepsilon_{\text{residual}}$ | $R^2_{\text{RLMM}}$ |
|---|---|---|---|---|---|---|
| $L_{A,eq,4h}$ | 51.7 [49.46,53.93] | 0.05 [-0.01,0.11] | 2.3, 0.13 | 1.08 | 2.86 | 0.15 |
| $L_{A10,4h}$ | 54.99 [52.66,57.32] | 0.05 [-0.01,0.11] | 2.75, 0.10 | 0.93 | 3.08 | 0.12 |
| **$L_{A50,4h}$** | **43.99 [41.84,46.15]** | **0.08 [0.03,0.13]** | **6.93, <0.01** | **1.12** | **2.71** | **0.25** |
| **$L_{A90,4h}$** | **28.98 [26.84,31.12]** | **0.08 [0.03,0.13]** | **7.91, <0.01** | **1.14** | **2.69** | **0.27** |
| *ONI* | 90.21 [87.30,93.12] | 0.03 [-0.04,0.10] | 0.12, 0.72 | 1.75 | 3.54 | 0.20 |
| *NPL* | 76.81 [74.19,79.43] | 0.03 [-0.03,0.10] | 0.23, 0.63 | 1.59 | 3.18 | 0.20 |
| ***NCl*** | **25.48 [24.69,26.27]** | **−0.02 [-0.04,-0.003]** | **5.79, <0.05** | **0.67** | **0.83** | **0.52** |
| ***PI*** | **13.04 [11.20,14.88]** | **−0.06 [-0.10,-0.01]** | **5.5, <0.05** | **0.28** | **2.68** | **0.13** |
| **$M_{A,eq}$** | **22.34 [21.47, 23.21]** | **−0.03 [-0.05,-0.01]** | **8.07, <0.01** | **0.65** | **0.99** | **0.48** |

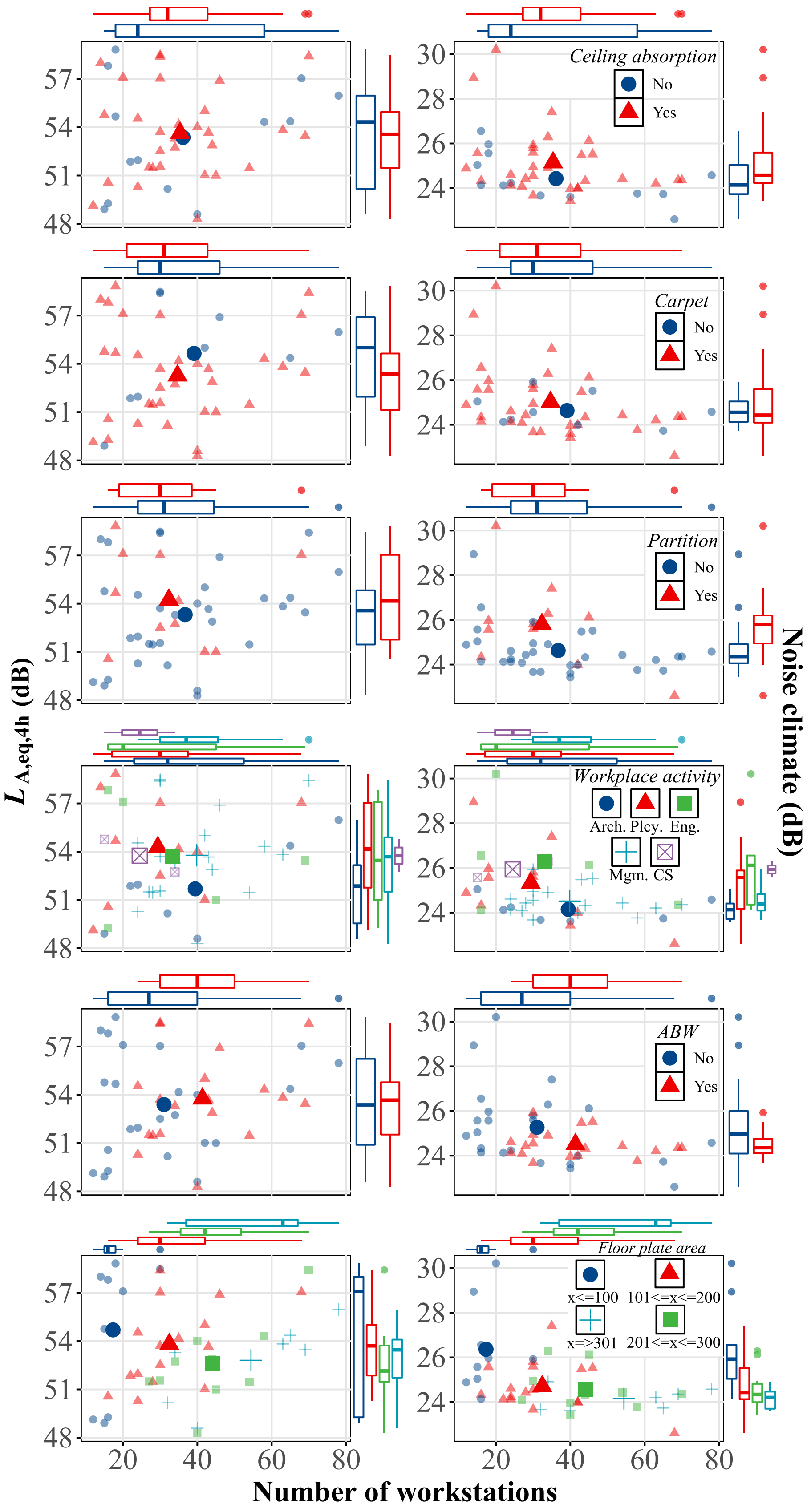

**Fig. 3.** Row-wise $L_{A,eq,4h}$ (left column) and *Noise climate* (*NCl*: $L_{A10,4h}$ - $L_{A90,4h}$ [8]; right column) values for key workplace parameters; refer to Table 1 for the abbreviations. Individual values and mean value (larger symbol) per group shown. Boxplots per grouping presented along the top (*x*-axis grouping) and right (*y*-axis grouping) margins of each plot.

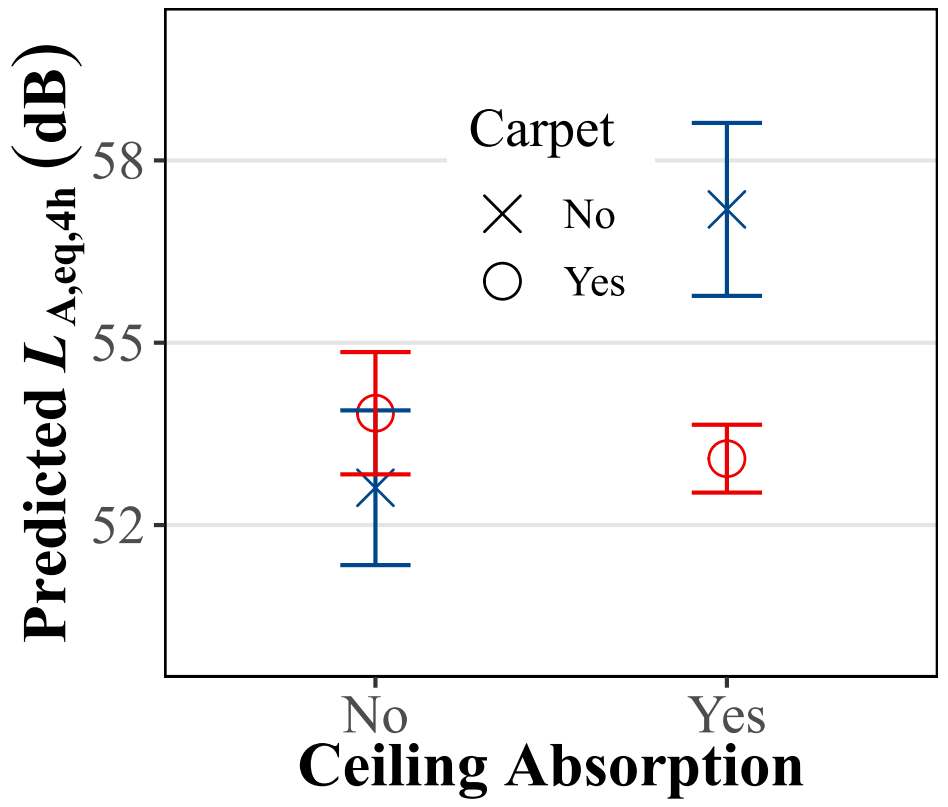

**Fig. 4.** The interaction of the Ceiling Absorption and Carpet groups in offices.

### 3.1.2. Based on one-third octave band results

Fig. 5 presents the one-third octave band levels for various groups per workplace parameter, and the overall levels for all groups combined.

### 3.1.3. Based on psychoacoustic metrics

Fig. 6 presents the mean value per office as a function of number of workstations ($WS_n$) for the psychoacoustic metrics (for loudness, $N_5$ following reporting recommendation in [37] and $N_{mean}$ provided). For these metrics, the effect of allowing the intercept per building to vary (random-effect), was not significant. Hence, results of GLM (i.e., no random-effect besides the residual error) for these metrics are presented in Table 4.

Compared to an intercept-only GLM, the number of workstations ($WS_n$) was not a significant predictor of *Loudness* (both $N_{mean}$ and $N_5$), *Roughness* ($R_{mean}$) and *Sharpness* ($S_{mean}$). However, the number of workstations ($WS_n$) was a significant predictor of fluctuation strength ($FS$) and loudness fluctuation ($N_{Fluctuation}$) in GLMs with each of these metrics as the dependent variable with $R^2$ values of 0.21 and 0.11, respectively. The stronger model with fluctuation strength as the dependent variable is presented in Eq. (4).

$$FS = 0.83 - 0.003 \times WS_n + 0.10_r \qquad (4)$$

### 3.2. For the omnidirectional measurements

Each boxplot in the left panel of Fig. 7 presents the daily $L_{A,eq,8h}$ values over an entire working week from 09:00–17:00 for offices F.24–F.30. The mean values ranged from 52.7 dB (Tuesday) – 51.1 dB (Monday) over the week, i.e., around 1.6 dB. The median values ranged from 52.5 (Wednesday) – 51.5 dB (Monday), i.e. around 1 dB. Daily mean values were within 0.5 dB of the respective median values, except for Wednesday where the mean was around 1 dB less than the median value, resulting from a positively skewed distribution.

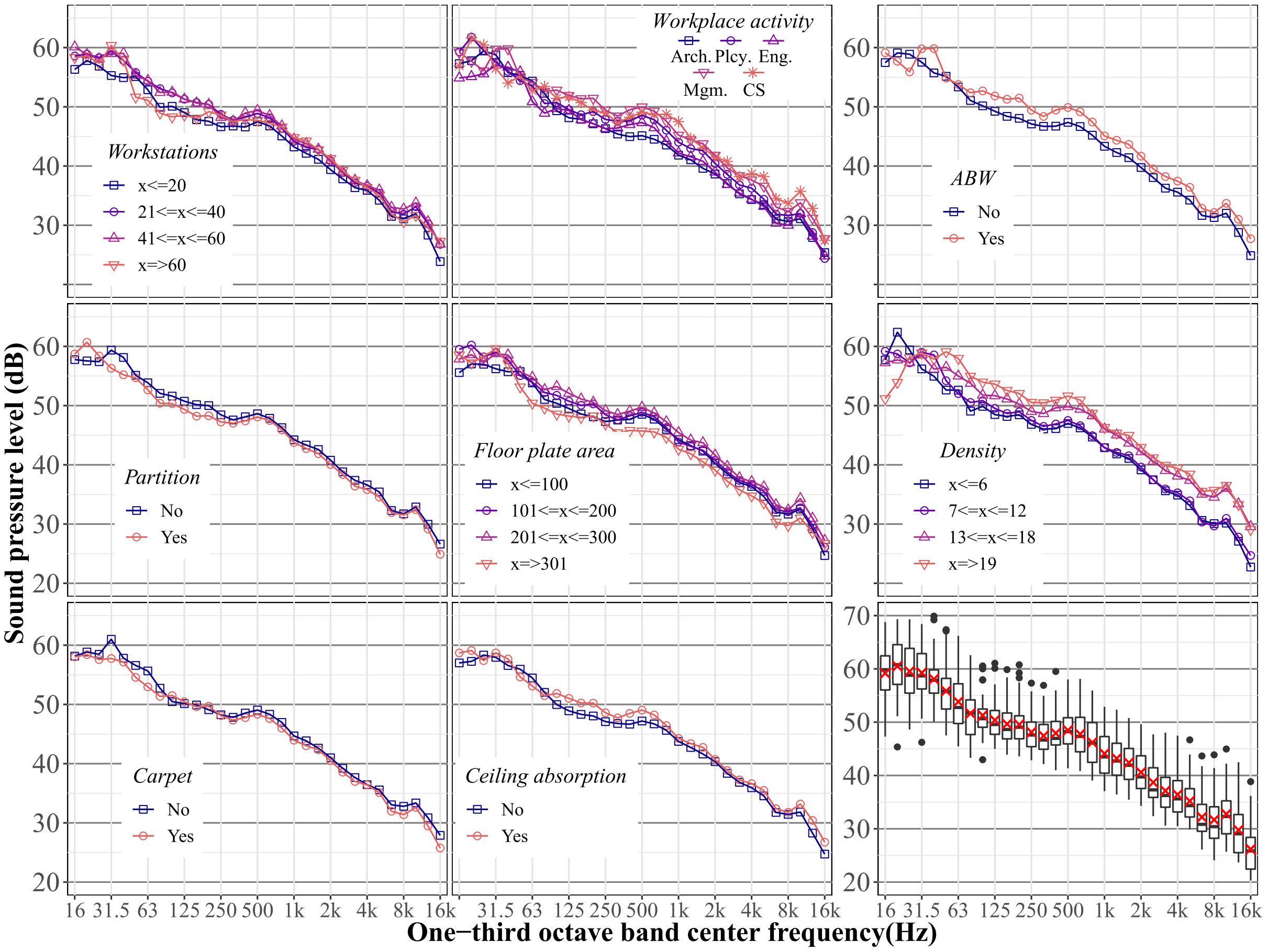

**Fig. 5.** One-third octave band levels grouped according to various workplace parameters. The bottom-right panel presents the overall results as box-plots per one-third octave band and the respective mean value as crosses.

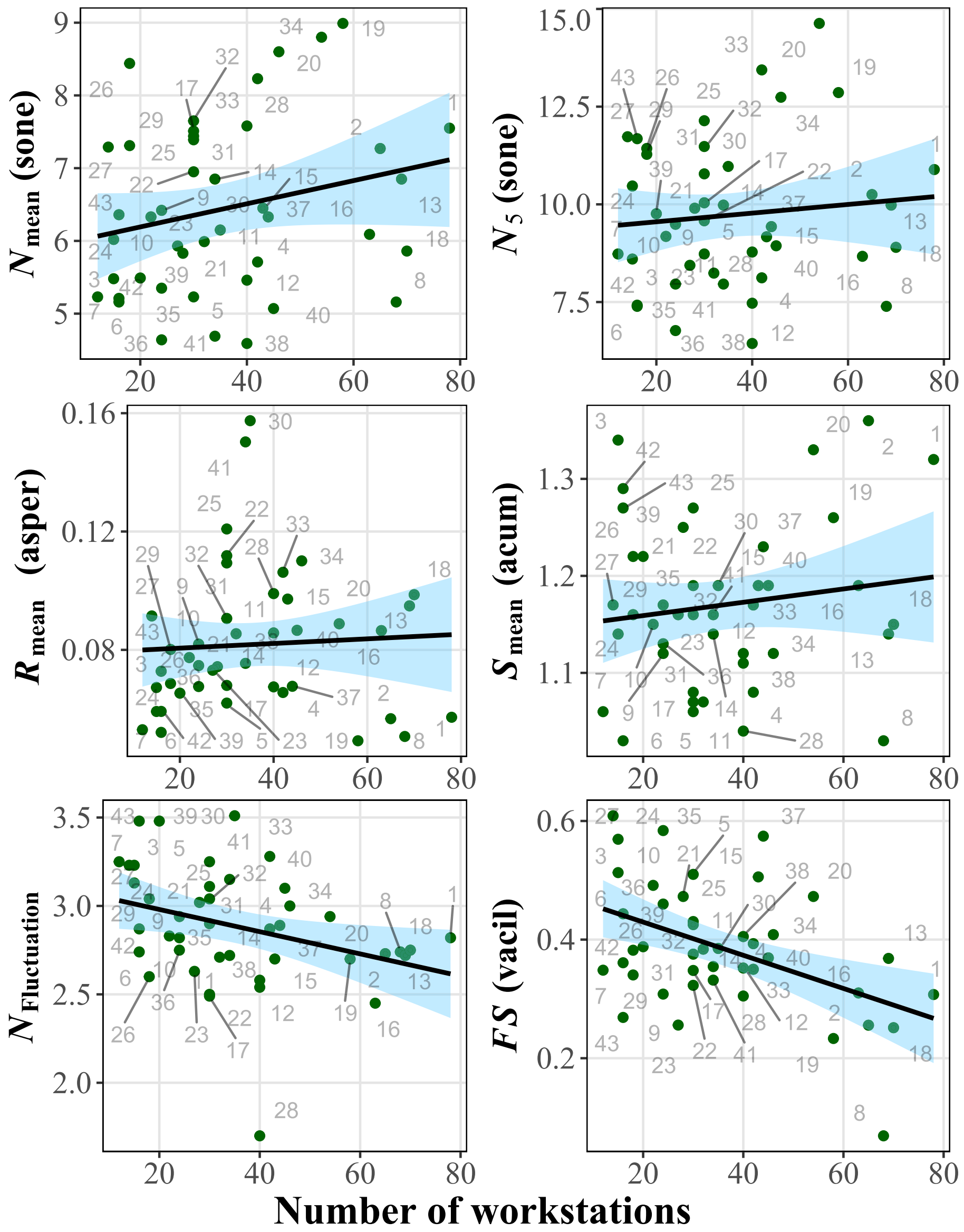

**Fig. 6.** Scatterplot per psychoacoustic metric, where each data point is labelled with its office number and presents averaged left and right ear values for all metrics except loudness, for which binaural summation is based on [26]. The GLM fit (Table 4) presented as dashed lines and shaded 95% confidence interval. *N*: Loudness, *R*: Roughness, *S*: Sharpness, *FS*: Fluctuation Strength.

**Table 4**
GLM parameters for the psychoacoustic metrics. See Table 3 caption for details about columns 2–5. Column 6 lists the goodness-of-fit of the linear model ($R^2$).

| *y* | ***a*** **[95% CI]** | ***b*** **[95% CI]** | $\chi^2_{\text{null vs. x}}(1)$, ***p-value*** | $\varepsilon_{residual}$ | $R^2$ |
|---|---|---|---|---|---|
| $N_{mean}$ | 5.87 [5.08,6.67] | 0.02 [-0.004,0.04] | 2.44, 0.12 | 1.17 | 0.06 |
| $N_5$ | 9.33 [8.08,10.60] | 0.02 [-0.02,0.04] | 0.47, 0.49 | 1.89 | 0.01 |
| $R_{mean}$ | 0.08 [-0.06,0.10] | $8\cdot10^{-5}$ [$-3\cdot10^{4}$,$5\cdot10^{-4}$] | 2.75, 0.10 | 0.02 | 0.01 |
| $S_{mean}$ | 1.15 [1.09,1.20] | $7\cdot10^{-4}$ [$-8\cdot10^{-4}$,$2\cdot10^{-3}$] | 0.86, 0.35 | 0.10 | 0.02 |
| $\boldsymbol{N_{Fluctuation}}$ | **3.11 [2.89,3.32]** | **$-6\cdot10^{-3}$ [-0.01,$-8\cdot10^{-4}$]** | **5.04, <0.05** | **0.32** | **0.11** |
| ***FS*** | **0.49 [0.42,0.55]** | **$-3\cdot10^{-3}$ [$-4\cdot10^{-3}$,$-1\cdot10^{-3}$]** | **10.30, <0.01** | **0.10** | **0.21** |

Each boxplot in the right panel in Fig. 7 presents the hourly $L_{A,eq,1h}$ values for offices F.24–H.41, which were measured using several sound level meters (SLM) per office over at least an entire day. The mean hourly values ranged from 51.2 dB (16:00–17:00) – 52.9 dB (10:00–11:00) in these offices, i.e., around 1.7 dB. Most hourly mean and median values show close agreement, differing at most by 0.4 dB (for 15:00–16:00). The overall mean $L_{A,eq,8h}$ was 52.1 dB, the median 52.2 dB, and a range of 43.0–61.2 dB (last boxplot in Fig. 7). The mean $L_{A,eq,4h}$ was 54.8 dB for these offices using the dummy head (Section 3.1), hence a difference of 2.7 dB.

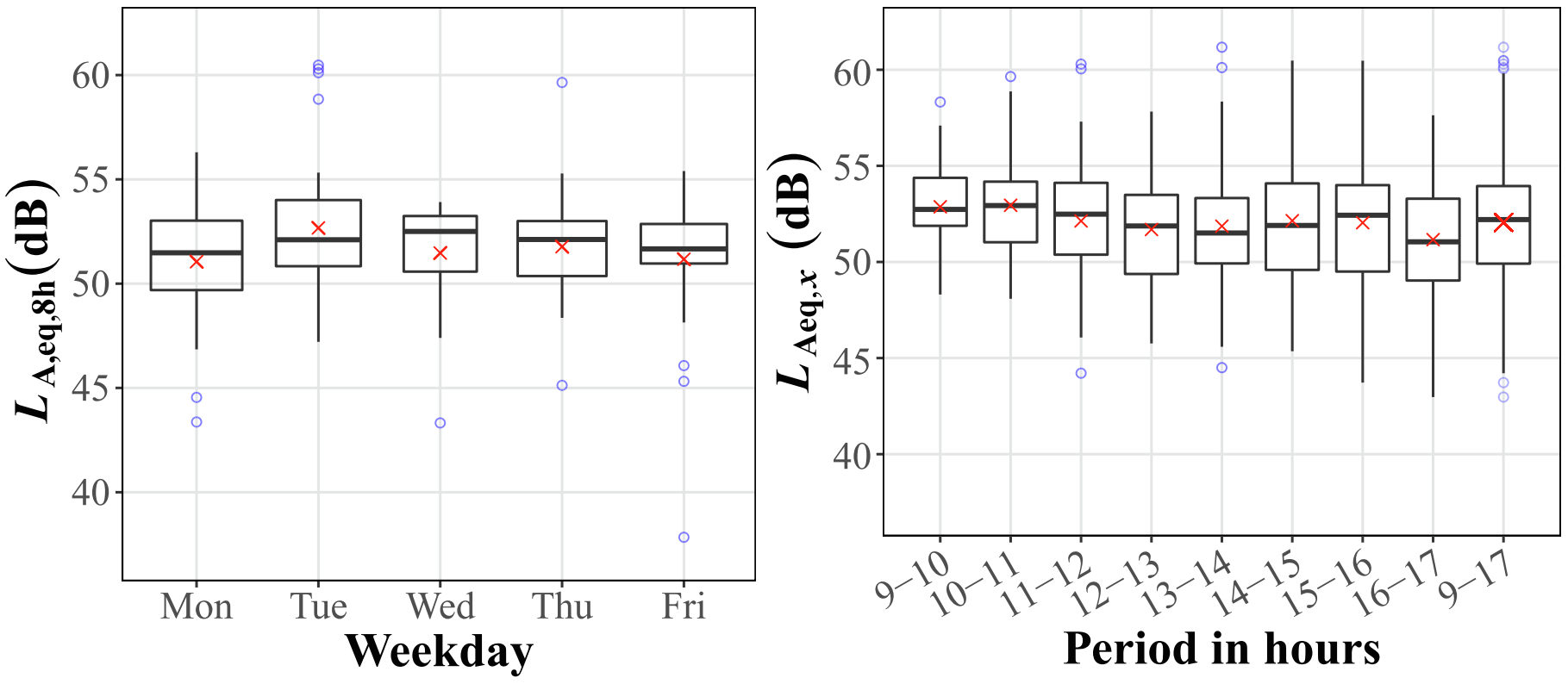

**Fig. 7.** Daily $L_{A,eq,8h}$ values using sound measurements over one week (left panel; offices F.24–F.30) and $L_{A,eq,x}$ values (right panel) for offices F.24–H.41 (18 total), where × is 1 h (1 h) for all but the last boxplot, where it is 8 h (8 h).

*3.3. Reverberation times in offices*

Fig. 8 presents the reverberation times ($T_{30}$) in the offices as a function of the number of workstations, where the individual offices are annotated alongside the values, and are grouped according to three ceiling height groups and two ceiling absorption groups. The values seem reasonable except the relatively high $T_{30}$ value of Office C.8, which can be attributed to the complicated ceiling design comprising mostly of hard surfaces. The right panel in Fig. 8 presents the binaural $L_{A,eq}$ values in offices as a function of $T_{30}$ values, where a wide variation in $L_{A,eq}$ values can be seen per $T_{30}$ value groups.

## 4. Discussion

In the following, Section 4.1 provides an historical review of SPL measurements in occupied OPOs, culminating in the current measurements. Sections 4.2–4.4 discuss the three main categories of metrics considered here: binaural levels, spectrally-defined metrics, and psychoacoustic metrics. The predictive models from these categories of metrics are compared in Section 4.5 and the study's limitations are discussed in Section 4.6.

*4.1. History of broadband sound pressure levels over 50 years of research in open-plan offices*

Appendix A provides an extensive summary of previous studies referred to in the following. Fig. 9 collates the mean and range of SPL values in OPOs reported by the current and eight previous studies that span 50 years and offices in at least five different countries. While the measurement method varies between these studies, they were selected because they provide results for multiple offices. The $L_{A90}$ chart the background noise trends due to HVAC and other machinery that is mostly active throughout, whereas the $L_{A10}$ values represent the noise peaks. Studies from 2000s report values for longer averaging periods, generally an entire working day, compared to studies from previous years. Yet, the authors of earlier studies reported what they considered representative samples, and hence, the findings from such studies can at least be compared for broader trends. This is supported further by the current data where the mean $L_{A,eq}$ (binaural) changed very little from 53.2 to 53.7 dB ($L_{A,eq,4h}$ = 56.3 dB) for averaging time intervals ranging from 5 min to 4 h, and the mean $L_{A,eq}$ (omnidirectional) varying by 1.6 dB over a working week and 1.7 dB over the working hours in a subset of offices (Section 3.2).

The earliest study from 1970 [4] in clerical OPOs reported the highest $L_{A,eq}$ in Fig. 9, within an extensive range of offices. The 1972 study was also quite elaborate in its scope, providing $L_{A10}$ and $L_{A90}$ values that are considered baseline values in the following. Compared to the 1970 study, the study from 1973 [4] reported considerably lower $L_{A,eq}$ values, although the quietest office included employees mainly performing concentrated work, where minimal conversations can be assumed. However, the highest value reported is still well below the average $L_{A,eq}$ of 64.4 dB reported in the 1970 study [4].

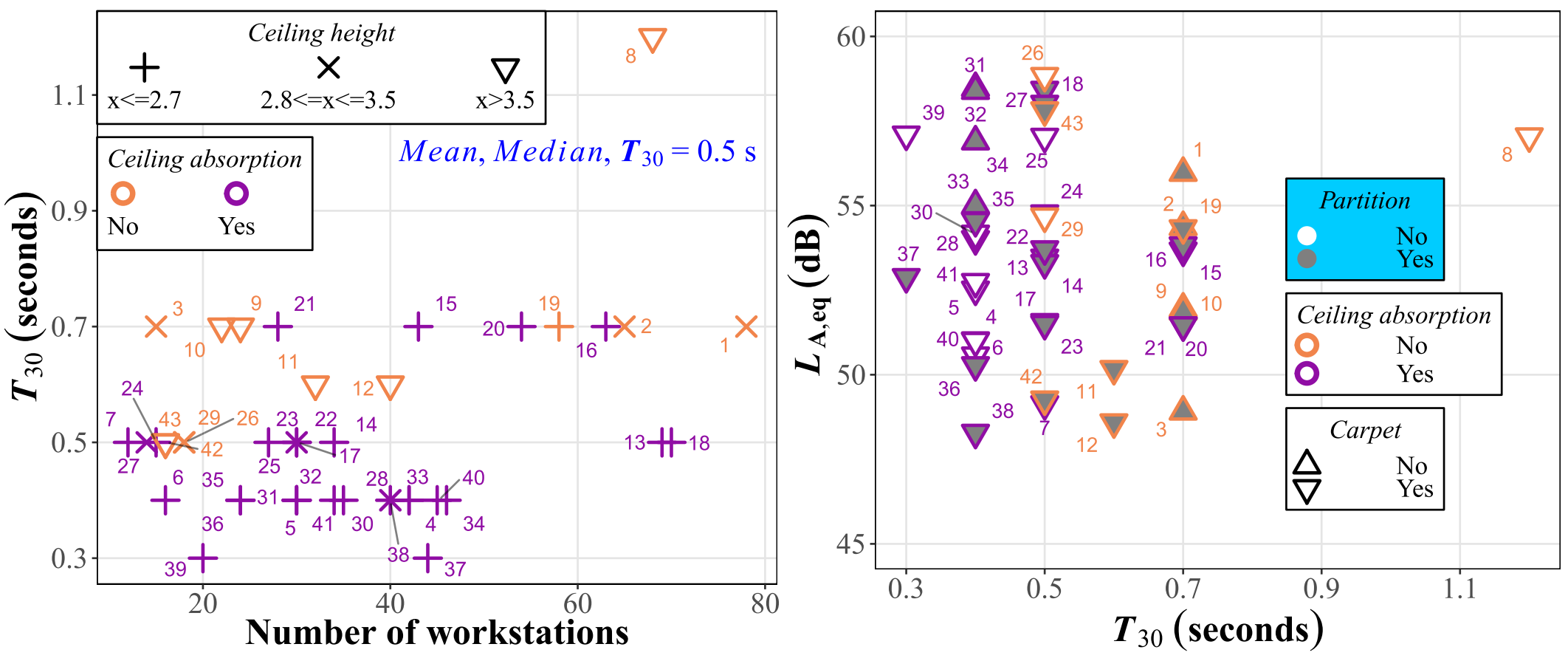

**Fig. 8.** (Left panel) Reverberation times ($T_{30}$) in unoccupied offices. (Right panel) $L_{A,eq}$ in offices as a function of $T_{30}$.

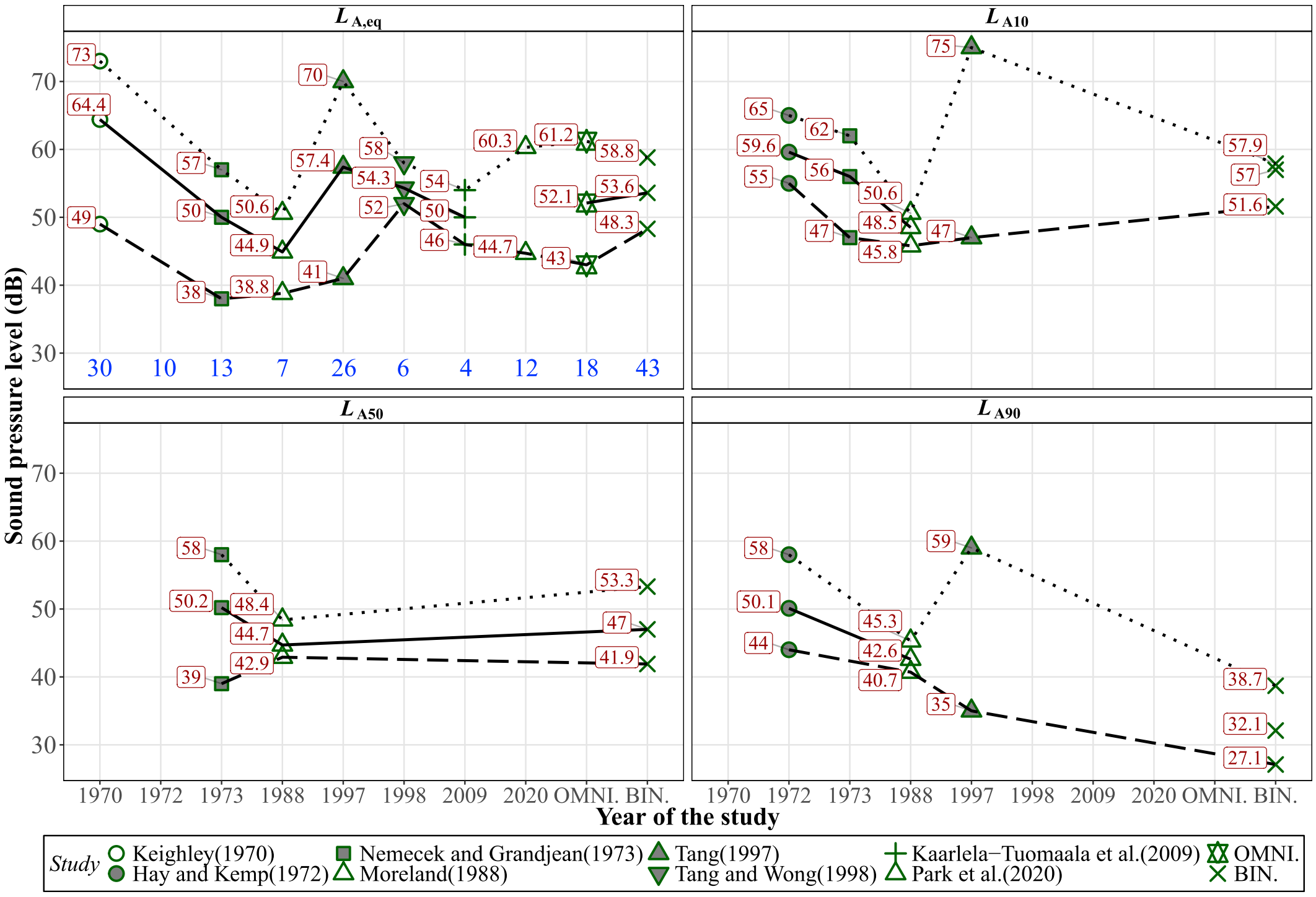

**Fig. 9.** Highest (connected by dotted line), Mean (connected by solid line) and Lowest (connected by broken line) values reported for omnidirectional measurements in offices by some previous studies and the omnidirectional (OMNI.) and binaural (BIN.) measurements in the current study. Landscaped OPOs, when reported, are denoted by filled data points. The number of offices per study is listed near the bottom of the top-left plot.

The $L_{A10}$ values in the 1973 study were lower compared to the 1972 study, and this downward trend continued in the 1988 study; the latter also showing a downward trend in the $L_{A50}$ and $L_{A90}$ values, and small ranges in the values of all metrics overall. This could partly be explained due to the 1998 study being conducted in cubicle-style offices (in the USA) with 1.5 m high screens, and hence higher absorption area, compared to the 70′s studies that were all conducted in landscaped (except perhaps [4]) offices (in Europe) with limited to no partitions.

The study from 1997 [10] was the first with offices in Asia, conducted in landscaped offices with dominant HVAC noise, reported relatively low $L_{A,eq}$ for the quietest office (41 dB; Fig. 9), but with mean and highest $L_{A,eq}$ substantially higher than the 1973 study, and closer to the 1970 study. A 1998 study by the 1997 study's group reported a smaller range of $L_{A,eq}$ values (6 dB [11] compared to 29 dB [10]), with less dominant HVAC noise than the 1997 study, and more contributions from office noise components (conversations, photocopiers, etc.), albeit within a much smaller sample. With lower HVAC noise, the highest $L_{A,eq}$ in the 1998 study is similar to the 1973 study (also with landscaped offices), and with dominant HVAC noise, the highest $L_{A,eq}$ reported is closer to the 1970 study. In the 1997 study, the highest $L_{A10}$ value reported is the highest overall, and the $L_{A90}$ value is substantially higher than the 1988 study and is similar to the 1972 study that, interestingly, also reported HVAC noise being dominant. All this information seems to suggest high SPL due to HVAC and intermittent high-impact noise from machinery etc. in the 1972 and 1997 studies. The lowest $L_{A90}$ value in the 1997 study, however, continues a downward trend compared to the previous years.

It is likely that with high $L_{A90}$ values due to HVAC noise, the conversations in some offices in the 1997 study might have involved the Lombard effect. Lombard effect has been shown to initiate around 43.3 dB to 45 dB $L_{A,eq}$ for broadband noise in some studies [38]. This is somewhat supported by the average SPL around the 500 Hz octave band (where a substantial portion of speech energy is concentrated) in the 1997 study being higher in comparison to the 1998 study (see Fig. 10), and slopes for the 500 Hz–2000 Hz octave bands being lower by around 1.0 dB, i.e., lesser decrease in SPL values for these bands.

The study from 2009 is the smallest sample of offices in Fig. 9 but has been cited in many recent studies to represent typical $L_{A,eq}$ values in offices (e.g., [39]). This study provides most of the relevant physical details about the sampled offices, the sound environment, with the measurements lasting a working day (7 h). The maximum $L_{A,eq}$ in this study is notably the lowest among the ones in Fig. 9, with the average $L_{A,eq,7h}$ being similar to the 1973 study. While both the 1973 and 2009 studies are conducted in European offices, the former has landscaped and the latter has cubicle-types offices, which may partly account for the similar $L_{A,eq}$ values despite machinery and HVAC noise presumably being lower in the 2009 study.

A study from 2020 [2] (listed in Appendix A, not in Fig. 9) reported a mean $L_{A,eq,7.30h}$ value of 53.7 dB from measurements in one OPO with a similar number of employees per office as in the 2009 study but with different room acoustics, among other factors. Both the 2009 and 2020 study are closer in scope to the current omnidirectional measurements since they sampled close to, and over many, workstation locations over more than 7 h. Another study from 2020 is also close in scope to the current set of omnidirectional measurements, wherein 12 OPOs were measured over 8 h each, with 9 offices sampled at a single workstation and 3 at three workstations [3]. The 15.6 dB range in the mean $L_{A,eq,8h}$ values in this 2020 study is 2.6 dB less than the 18.2 dB range in the current study with a larger sample of offices (Figs. 10 and 7). This 2.6 dB difference may be considered reasonably similar given the methodological, room acoustic, and work-culture differences between the workplaces in these studies.

Other notable studies not listed in Fig. 9 but listed in Appendix A from the 2000–2020 period include a 2003 study reporting mean

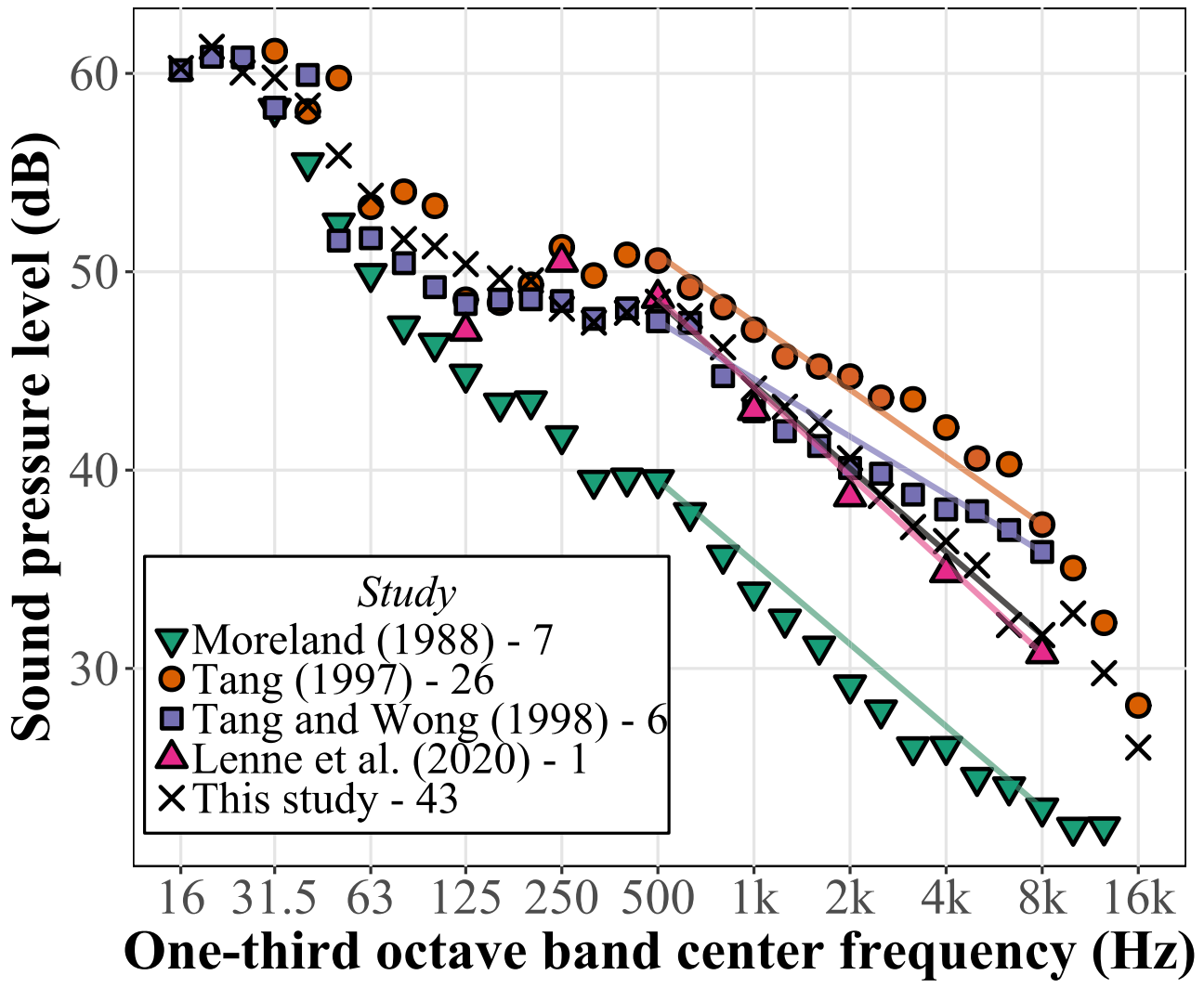

**Fig. 10.** One-third octave band levels in three previous studies and the current, with the number of offices per study listed in the legend. The slope between 500 Hz and 8000 Hz octave bands shown per study. The one-third octave band values for Lenne et al. [2] are calculated by subtracting 4.8 dB ($10 \times log_{10}(3)$) from the octave band values reported in their study, assuming octave band powers to be equally distributed between constituent one-third octave bands.

$L_{A,eq,20min}$ of 55.1 dB over several measurements and a range of 45.8–62.6 dB [40], and a 2005 study reporting mean $L_{A,eq,5m}$ values over several measurements and locations in two offices as 55 and 60 dB. These studies, however, do not provide as many physical details about the offices or the measurement method as [41] and [2].

Taken together, studies over the last 50 years, including more recent ones from the 2000′s, report a fairly wide range of $L_{A,eq}$ values, which should be considered carefully within laboratory studies that need to decide on representative levels for offices, in policy documents (cf. [42,43]), etc. The $L_{A10}$ and $L_{A50}$ values show a downward trend over the years, although not always. Since only a few studies have reported these percentile levels, the downward trend in values is harder to justify. However, *the lowering of $L_{A90}$ values shows a more consistent trend over the years, which largely denotes lowering background noise due to mostly quieter office machinery (including HVAC) over the years.*

### 4.2. Perspectives of binaural levels in open-plan offices

While the discussion based around omnidirectional measurements in Section 4.1 is useful, the binaural levels presented in Figs. 1–3 are more representative overall since they are based on a larger sample (43 offices), and present values that incorporate some of the acoustic effects of a human head, including pinna and ear canal characteristics. This not only enables a closer approximation of hearing levels in offices in general but is also useful in establishing and determining more realistic levels for policy documents (e.g., [42,43]), laboratory experiments, etc. In terms of policy documents, a recent French standard (NF S 31–199:2016) recommends four target values depending on the prominent work activities in OPOs: $48 < L_{A,eq} < 52$ dB for workplaces where main activities are performed over telephones, $45 < L_{A,eq} < 50$ dB for workplaces where prominent activities include spoken communication in-person and over telephones for collaborations between employees over telephones, $40 < L_{A,eq} < 45$ dB for workplaces where prominent activities involve individual work requiring concentration and sporadic and quiet conversations, and $L_{A,eq} < 55$ dB for reception areas in workplaces [42]. The workplaces in the current paper include all these categories except the latter (reception areas), although a strict division between the remaining three workplace groups according to the French standard is not possible (or intended within the standard) for the current sample of offices since they usually consisted of groups of occupants performing various activities. Nevertheless, the mean $L_{A,eq}$ of 53.6 dB for the current sample of offices is reasonable only for workplaces involving regular conversations according to this standard and excessive for all other workplace groups in general, and also for the five predominant workplace activity groups identified in the current study, as seen in Fig. 3 (top-left panel).

For experiments in cognitive psychology, etc., the summary statistics provided in Figs. 1–2, and spectral information provided in Fig. 5 are recommended for determining sound reproduction levels for a wide range of experimental trial durations. The applicability to several trial durations is reasonable given the maximum $L_{A,eq}$ change for a large variation in averaging durations for the current sample was quite small (around 0.5 dB).

For most of the relevant workplace parameters considered, there was little variation in the $L_{A,eq}$ values overall, as seen in Fig. 2 and based on further statistical analyses (Section 3.1.1). Hence, while the $L_{A,eq}$ values for the entire sample (Fig. 1) can represent values for the relevant categories of these workplace parameters, it might be more useful to use $L_{A,eq}$ values per category in certain cases (Fig. 2). However, some of these findings need to be considered in relation to the disproportionate sample sizes in certain categories (see Table 1, Fig. 2 and Section 3.1.1). Specifically, in relation to workstation partitions, most offices (74.4%) including many non-ABW offices did not have partitions, typifying the trend in more modern fit-outs with low-rise partitions or none or all, compared to the high-partitions trend from the 1970 s to early 2000 s [1,12]; only 5% of the offices with predominantly customer service focus, although almost all offices were multi-department workplaces; and offices with floor plate areas between 101–300 m$^2$ constituting almost 63% of the current sample.

Furthermore, ceiling absorption as an individual factor did not lead to significant differences in the $L_{A,eq}$ values (Section 3.1, Fig. 4). The role of ceiling absorption has consistently been shown to be one of the most important in controlling the room acoustics of offices [44–46]. Most notably, a recent study by Keränen et al. tested an extensive set of sound absorption profiles for ceilings, walls, and partitions in offices. The floor had non-absorptive vinyl covering throughout, and the testing was performed in a 9.41 × 8.94 × 2.55 m room. In their study, the presence of ceiling absorption had the biggest impact on the relevant ISO 3382–3 metrics ($D_{2,S}$ and $L_{p,A,S,4m}$) and $L_{p,A,S,2m}$; the latter consistent with two previous studies where a 2 m distance was used to represent nearby workstations [44,45]. The introduction of partition screens increased the positive effect of ceiling absorption, while wall absorption had little or even a slightly detrimental effect in some cases (see [46] for details). However, Fig. 4 shows that the predicted $L_{A,eq}$ was 4.1 dB lower in the current sample of offices with full-floor carpeting than ones without (also statistically significant), i.e., presence of carpeting was important individually and in combination with ceiling absorption. Typical textile carpeting does not increase sound absorption by much, and has been shown to not appreciably affect the level of loudspeaker reproduced speech in a laboratory study [45]. Hence, to explain the surprising finding regarding the effect of carpeting in the current study, it is hypothesised that since carpeting reduces noise due to footfalls and other floor impact sounds, this leads to lowered speech levels in offices due to a psychological process similar to, or indeed, the Lombard effect.

Furthermore, the right panel of Fig. 8 shows that $L_{A,eq}$ values in offices were mostly independent of respective reverberation times,

and hence, independent of the apparent sound absorption in the room if diffuse fields are assumed in offices; while arguable, such assumptions were reasonable within empirical findings in [46] for several absorptive profiles of the testing room that is smaller than all offices in the current sample. The absence of relationship between $L_{A,eq}$ and room absorption could also partly explain why there was no relationship between $D_{2,S}$ and % of highly annoyed participants in a recent extensive study in OPOs [47]. However, it can be argued that the current sample of offices is somewhat lacking in terms of proportions of absorptive profiles of ceilings, partitions, etc., and does not have offices with sound masking systems (Table 1). Overall, the role of absorptive treatment (ceilings, partitions, etc.) in achieving suitable room acoustics is indubitable in unoccupied offices (cf. [1] for a historical trend in OPO designs), and sound masking may play a role too (see [2] for a review). The contribution of the current findings is in showing that actual sound levels during occupancy are influenced by non-acoustic workplace factors and may be quite complicated to predict based solely on room acoustic data. In order to test this systematically, *an ambitious set of studies are proposed here, wherein the absorptive profiles of surfaces and sound masking within offices are varied similar to Keränen et al. [46], albeit in actual OPOs and/or realistic simulations, to examine combined effects on the occupied sound levels. Herein, phenomena such as Lombard effect due to carpeting, as hypothesised above, can be directly measured.*

### 4.3. One-third octave band spectra and spectrally-defined metrics in offices

Fig. 5 shows relatively higher SPL variations between most categories per workplace parameter for center frequencies below 63 Hz; a negative slope starting from around 63 Hz with relatively lesser yet still noticeable SPL variations for 63 Hz–500 Hz center frequency bands, with a local peak around 500 Hz; and a steeper slope between 500 Hz and 8000 Hz. There is a local maximum around 10 kHz (Fig. 10) which is partly due to the dummy head microphones' sensitivities in this frequency region. The average slope between 500 and 8000 Hz (Fig. 5, last panel, and Fig. 10) octave band center frequencies is −4.4 dB/octave based on median SPL values, and −4.0 dB/octave based on mean SPL values; −3.4 dB for 500–1000 Hz, −3.9 dB for 1000–2000 Hz, −4.5 dB for 2000–4000 Hz, and −4.1 dB for 4000 Hz–8000 Hz octave bands. The average slope between 31.5 Hz and 16000 Hz is −3.7 dB/octave based on mean SPL values.

For the various workplace parameters (Partitions, Carpeting, etc., Fig. 5), variations between respective categories in the 500 Hz and 2000 Hz band are considered here. Along with their respective adjacent bands, the former band is important since most of the speech energy will be concentrated in the region, and the latter band is important for speech intelligibility with lesser SPL considered better (among other factors). The most notable differences emerged between offices characterised by different workplace activities, with Architecture/Design workplaces having around 5 dB lower SPL than the Management workplaces for the 500 Hz band – implying reduced speech-related activity in the former – and around 3 dB lower SPL than the Management and Customer Service workplaces for the 2000 Hz band. Speech is also likely to be slightly more intelligible in the Architecture and Design offices than the Management and Customer Service offices. Fig. 3 also shows lowest $L_{A,eq,4h}$ values for Architecture and Design workplaces. The ABW offices had around 2.5 dB and 2 dB greater SPL in the 500 Hz and 2000 Hz bands, respectively, than the non-ABW offices. For the four floor plate area categories, the largest offices in the current sample had lower SPL values than the other three categories which varied by around 1 dB between them. Offices with floor plate areas greater than 300 m$^2$ were around 3 dB lower in the measured SPL than the other categories of floor plate areas for the 500 Hz band, and around 2.5 dB lower SPL than the offices between 201–300 m$^2$ for the 2000 Hz band. For workstation densities (number of workstations per 100 m$^2$) there were two distinct groups: the two lowest density categories were around 4 dB and 3.5 dB lower in the measured SPL for the 500 Hz and 2000 Hz band, respectively, compared to the two highest density categories. The rest of the workplace parameters – Carpeting, Partitions, Ceiling absorption categories – had less than 2 dB and typically much lower differences between their respective categories (Fig. 5).

In comparison (Fig. 10), Moreland [12] reported a −4.0 dB/octave slope for the 500 Hz–8000 Hz octave bands, which is the same as the current study, and a −4.0 dB/octave slope for 31.5 Hz–8000 Hz octave bands, which is −3.4 dB/octave for the current study. However, for the 500 Hz–8000 Hz band, the SPL values in Moreland [12] are on average 9.5 dB lower than the current study; lower overall beyond the 63 Hz band than other studies in Fig. 10; with limited evidence of much speech-based contribution due to the relatively flat area around the 500 Hz band; and a relatively smooth broadband slope. Two studies in Fig. 10 that were conducted in landscaped offices with presumably lower absorptive treatment (among other factors) reported shallower slopes than the current study: Tang [10] reported −3.1 dB/octave and Tang and Wong [11] reported −3.3 dB/octave slopes. Lenne et al. [2] reported −4.5 dB/octave slope from measurements in one office.

For acceptable HVAC noise in offices, a −5 dB/octave slope is often quoted as the optimum or 'neutral' [48] following [49], although it has been disputed by other studies [23]. Pseudorandom noise with −5 dB/octave slope is also common in sound masking systems in OPOs [2,50]. Overall, the slopes in the current study and previous studies listed in Fig. 10 (except perhaps Moreland [12]) report an obvious influence of speech energy in comparison to the more 'neutral' HVAC slopes. *Based on the current results, a comparison of pseudorandom masking noise with a slope of −4 dB/octave* vs. *−5 dB/octave (indeed other slopes, and/or spectral adjustments similar to [24]) at various overall levels can be recommended, to comprehensively determine the masking effect of such spectra against speech in OPOs.*

### 4.4. Psychoacoustic metrics

Psychoacoustic metrics such as loudness ($N$), roughness ($R$) have been used extensively for sound perception evaluation in a variety of contexts such as speech [51,52], HVAC noise [53,54], road traffic noise [55], refrigerator noise [56], etc. The use of such metrics in OPOs has been limited in comparison. Major studies include Tang [10] where loudness and loudness levels (in phons) in occupied offices were considered to explain occupants' subjective impressions, and Schlittmeier et al. [57] where the so-called irrelevant speech effect (ISE; decline in short-term memory performance due to task-irrelevant background sounds) was modelled as a function of fluctuation strength ($FS$), based on laboratory-based ISE experiments using speech and non-speech sounds including office sounds [57].

From a novelty perspective, the current paper is the first to report key psychoacoustic metrics using long-term binaural measurements in a large sample of open-plan offices. In that regard, the loudness values reported in 43 offices here provide more reliable representations for both summary and individual values across a wide variety of contemporary offices than those reported for the 26 landscaped offices in Tang [10] in which HVAC-generated noise predominated. The values reported in Tang [10] are not directly comparable to the current ones because they were based on monaural recordings (cf. binaural in the current study), and because they were based on a different loudness evaluation method (Zwicker's method [58] as implemented in [59]) to the non-standardized binaural loudness [26] method used in the current study.

The mean *FS* over all the offices was 0.39 vacil, which is reasonably close to the *FS* values for recordings of real office noise (0.41 vacil for three offices and 0.46 vacil for another office recorded using a binaural dummy head [60]) used in Schlittmeier et al. [57]. Using their predictive model [57], the median ISE in the current study for mean *FS* of 0.39 vacil is 4.3%, i.e., the median performance detriment predicted in verbal short-term memory recall task relative to silence condition is 4.3%. This value is well within the 95% CI of participants' performance in short-term memory tasks for the office sound stimuli in [57].

It can be argued that *FS* alone is not a comprehensive predictor of ISE in offices without additional considerations regarding semanticity of speech, etc.; see discussions in [57,61] and compare another ISE model based on speech transmission index in unoccupied offices [62–64]. Besides, ISE represents just one of the many aspects of auditory distraction in offices. One of the major limitations about psychoacoustic metrics tends to be the relatively higher computational costs in their calculation relative to the ubiquitous level-based metrics, which have also demonstrated to be better predictors of subjective aspects of OPO acoustics [6,11]. Yet, *the current FS data, and indeed observations on other psychoacoustic metrics across a wide variety of offices, presents opportunities to develop and test data-driven hypotheses regarding the role of these metrics in perceptual assessments of offices in-situ and especially in controlled laboratory experiments.*

### 4.5. Comparison of predictive models of sound levels in OPOs

For effect size comparisons in the following, the guideline provided in [65] is followed to an extent, with $R^2$ values (for both GLMs and RLMMs) of 0.04, 0.25 and 0.64 taken as thresholds for small, medium and large effects. Since there is not enough context from previous studies, these thresholds are indicative only and may be refined with future work. Note also that the thresholds used here are stricter for the medium and large effects compared to the more commonly used thresholds suggested in [66].

In terms of overall trends, for the metrics based on A-weighted levels, Fig. 2 shows that values of $L_{A,eq}$ and the three percentile levels increased as the number of workstations ($WS_n$) increased. $WS_n$ significantly predicted the $L_{A50}$ and $L_{A90}$ values with small effect sizes, but not $L_{A,eq}$ and $L_{A10}$. While it is beyond the scope here, several studies over the years have reported $L_{A,eq}$ to be a significant predictor of some aspects of office occupants' subjective auditory impressions, with acceptability typically decreasing with rising $L_{A,eq}$ [3,6,10,11]. This suggests that although $L_{A,eq}$ may not exhibit a strong linear increase with increasing number of workstations, rising $L_{A,eq}$ values are still perceived negatively by office occupants. This is especially interesting as it leads to several research questions regarding the effects of the office size, workstation density, etc. of a workplace on its sound environment, as quantified using metrics characterising occupied (e.g., [3,11]) or unoccupied offices (e.g., [3,47]). In this regard, the current paper provides an extensive dataset to enable systematic studies comparing workplace parameters with its acoustics.

Out of the composite metrics based on A-weighted levels, $WS_n$ significantly predicted *NCl*, $M_{A,eq}$ and *PI* and the values of these parameters decreased overall with increasing number of workstations. The model predicting *NCl* as a function of $WS_n$ showed medium-large sized effect and is the largest overall out of all parameters considered in Table 2. While noise rating (*NR*) was significantly predicted by $WS_n$ (a medium-large sized effect; Eq. 4), it is not discussed further; see last paragraph in Section 4.3.

The trends for the psychoacoustic metrics are weaker in comparison in general (see Tables 3 and 4, and Fig. 6) with $WS_n$ not being a significant predictor of loudness, sharpness and roughness, and the latter two not showing a clear trend overall. For loudness (both $N_{mean}$ and $N_5$), consistent with the A-weighted levels, an overall increasing trend is seen with increasing $WS_n$ and a small effect size for $N_5$. Most notably, consistent with the level-based fluctuation metrics, both fluctuation of loudness ($N_{Fluctuation}$) and fluctuation strength (*FS*) show a decreasing trend with increasing $WS_n$ and the latter significantly predicts the former two with small effect sizes.

The models predicting *NCl* and *FS* values as a function of $WS_n$ are very interesting as they encapsulate the broad idea of fluctuating sound, while being quite different in their domains and calculations. *NCl* simply denotes broadband level fluctuations above the background noise in occupied offices ($L_{A90}$) whereas *FS* has a more constrained scope. *FS* characterises the subjective perception of amplitude modulations of up to 20 Hz with a maximum sensitivity at 4 Hz which corresponds to typical frequency of syllables in speech [67]. However, despite the contrast in sophistication, amongst other factors, the current results clearly show that more workstations/occupants lead to lower level-based fluctuations and fluctuation strength values. As discussed in Section 4.4, lowering *FS* values signify lowers short-term memory detriment, which essentially means that larger offices with more workstations are likely to have less ISE-based auditory distractions; this would need further testing. *Since NCl (or indeed other similar metrics such as* $M_{A,eq}$*) is much easier to calculate than FS, studying its relationship with auditory distraction in laboratory experiments and even in real offices is the main recommendation of this paper in this context.*

### 4.6. Limitations

The main limitation of this paper is that the data is not uniformly distributed across categories of several workplace parameters, mostly since the sampled office are limited to where necessary management approvals were received. For instance, most of the offices sampled did not have partitions, most offices were carpeted, and management-related work activity formed a large proportion compared to other work categories (Table 3). Future studies can focus on categories where the current dataset has uneven distributions of offices.

Offices with additional sound masking typically have $L_{A90}$ between 40 and 45 dBA (potentially higher in some countries), which is high compared to even the highest $L_{A90}$ of 39 dBA in the current sample (Figs. 1,9). Hence, another major limitation here is that it is not clear if the current results apply to such offices, and more studies are needed to clarify this.

The dummy head used in the paper does not incorporate the acoustic effect of shoulder and torso that is possible with head and torso simulators. Furthermore, the head position was always fixed per location, which does not incorporate the acoustic effect of head movements. These effects can be further explored in future studies.

## 5. Conclusions

This paper has examined the acoustic conditions in 43 occupied open-plan offices, in the context of prior comparable studies of occupied offices. Main conclusions are:

1) Over the last 50 years, the background noise in occupied offices has become quieter, as characterised by lower $L_{A90}$ values. $L_{A,eq}$, $L_{A10}$, and $L_{A50}$ show more complicated trends of values over the years but lean towards quieter offices overall.

2) Mean binaural $L_{A,eq}$ at workstations for the current sample is 53.6 dB with 95% CI: [52.7,54.5], which is likely to be representative of a typical working week, and is recommended in laboratory experiments for characterising open-plan office sound environments.
3) SPL in occupied offices follows an approximately −4 dB/octave decay over 31.5 Hz–16000 Hz octave bands compared to the −5 dB/octave slope typically assigned to HVAC operation in unoccupied offices.
4) Architecture and design offices in the current sample were the quietest overall and had lesser speech energy compared to other types of offices. Activity-based/flex-desk/non-territorial workplaces show some evidence of more speech energy compared to more traditional open-plan offices.
5) Metrics based on sound fluctuation show the most promise in terms of characterising variation in the office sound environment as the number of workstations increase, and larger offices are likely to have lesser auditory distraction as characterised by short-term memory performance. In this regard, a simple-to-calculate acoustic metric, the Noise Climate ($L_{A10}$ – $L_{A90}$), outperforms the computationally expensive psychoacoustic metric Fluctuation Strength.
6) An absorptive ceiling alone is not associated with reduced occupied sound levels in offices, but offices with absorptive ceiling *and* carpeted floor are quieter, suggesting that the reduced floor impact noise due to carpeting may lead to reduced speech levels via a process similar to the Lombard effect.

## Declaration of Competing Interest

The authors declare that they have no known competing financial interests or personal relationships that could have appeared to influence the work reported in this paper.

## Acknowledgements

This study was funded through the Australian Research Council's Discovery Projects scheme (Project: DP160103978) and the Deutsche Forschungsgemeinschaft (DFG, German Research Foundation) Research Grant scheme (Project number: 401278266). The authors thank Jonothan Holmes, James Love and Hugo Caldwell for assistance with the measurements.

## Appendix A

Summary of previous studies of sound measurements in occupied open-plan offices, following the terminology used in the source papers as closely as possible. Average and range presented when available.

| Study | Room type(s) | Ambient sounds | Method | Reported values |
|---|---|---|---|---|
| Keighley (1970) [4] | 30 offices out of 40 large offices in the UK reported here. These 30 offices were classified as (1) multiple function and general offices with some machines and (2) clerical offices with no machines. | Most workers performed clerical work, with lower proportions of typing pools, card-punch, business machine and machine rooms, and drawing offices. | Sound measurements on typical days (details not provided) at locations approximately midway between working positions. The most frequent $L_{A,eq}$ value over a 10 seconds interval taken as the average per reading. The mean over 2 days of such readings taken as the representative level. 1-minute tape recordings over 20 minutes taken over 2 days for further analysis. | $L_{A,eq}$ = 64.4 dB, Range: 49–73 dB |
| Hay and Kemp (1972) [5] | 10 mixed-personnel landscaped offices in the UK with full carpeting and acoustic ceilings. 2 offices with no HVAC not covered here. Offices ranging from 309–2212 $m^2$. | HVAC noise, People talking, telephones ringing and office machinery, ingress of external noise. HVAC noise dominant. | Offices divided into several areas. In each area, several locations measured using 60 seconds recordings every 20 minutes between 08:30–16:45 excluding lunch hours. | $L_{A10}$ = 59.6 dB; 55–65 dB$L_{A90}$ = 50.1 dB; 44–58 dB $T$ (500Hz) 0.3–0.8s. |
| Hoogendoorn (1973) [16] | 2 landscaped offices in Sweden, 900 and 1200 $m^2$ floor area with 1 cm thick carpet, absorbent ceilings. Occupation density ranged from 1 person every 11 $m^2$, and 1 person every 10 $m^2$. Acoustic partitions for about 1 out of 4 persons, ranging in height from 1.2–1.5 m. | Paperwork, telephone conversations, typing, calculators | Measurement details not provided | $L_{A,eq}$ = 52–58 dB, Range = 45–65 dB. |

(*continued on next page*)

**Conclusions** (*continued*)

| Study | Room type(s) | Ambient sounds | Method | Reported values |
|---|---|---|---|---|
| Nemecek and Grandjean (1973) [68] | 13 landscaped offices (2 that did not have HVAC not reported here) in 6 companies in Switzerland. Floor area between 252–1355 $m^2$ and 7.3–14.4 $m^2$/person. 9 offices with monothematic work, 4 mixed-function, with limited to no relation between work areas. | Conversations, office machines, telephone rings, movement back and forth, and traffic and industrial noise. | 3 sound measurements daily at 9–15 measuring points depending on room size, over 3–6 days. Each measurement lasted for 15 minutes and instantaneous dB(A) values recorded every 5 seconds. | $L_{A,eq}$ ~ 50 dB; 38 (office for concentrated work) –57 dB (punch-card dept., largest office)$L_{A50}$ = 50.2 dB; 39–58 dB$L_{A10}$ = 56 dB; 47–62 dB$L_{A1}$ = 59.7 dB; 51–65 dB |
| Boyce (1974) [69] | A large OPO in the UK. Further information not provided. | Unspecified | Unspecified | $L_{A,eq}$ = 54 dB, Peak = 62 dB |
| Moreland (1988) [12] | 7 carpeted OPOs with individual workstations (screen height 1.5 m; 5.6–7.4 $m^2$ floor area) of the cubicle type in the USA. Ceiling height: 2.7 m, with 0.6×1.2 m ceiling absorption panels suspended on a T-grid system. | Unspecified | Sound measurements at an empty workstation surrounded by other workstations in each office. 10 seconds measurements averaged over 2 minutes. Reported values for between 09:00–16:00. | $L_{A10}$ = 48.5 dB; 45.8–50.6 dB$L_{A50}$ = 44.7 dB; 42.9–48.4 dB $L_{A90}$ = 42.6; 40.7–45.3 dBOne-third octave band results. |
| Tang (1997) [10] | 26 landscaped offices in Hong Kong. Details not provided. | HVAC noise was dominant, not office noise (cf. [11]) | 5-minute sound recordings (no further details) | $L_{A,eq,5min}$ = 41–70 dB. $L_{A10}$ = 47–75 dB; $L_{A90}$ = 35–59 dB. See paper for other metrics. |
| Tang and Wong (1998) [11] | Follows up from [10]. Clerical work in 6 landscaped offices in Hong Kong. Geometry and size similar (not reported further). | Conversations, telephones ringing, laser printers and photocopiers | 5-minutes sound recordings close to the workers and at least 1 m from reflecting surfaces during normal operation hours. | $L_{A,eq,5min.}$ = 52–58 dB, without people 46–52 dB. See paper for other metrics. |
| Landström et al. (1998) [13] | Details not provided except that measurements were conducted in offices in industry and public administration sectors in Sweden. Not clear if offices were open-plan. | Fan noise, noise from office machines, telephone signals, conversations, and impact sounds of workers. | A 15-minutes representative sound recording at each of 71 office employees' workstations | $L_{A,eq,15min}$ = 53.3 dB, SD = 4.6 dB. $L_{Z,eq,15min}$ = 60.4 dB, SD = 3.6 dB. One-third octave band results. |
| Leather et al. (2003) [40] | Office of a government finance department in the UK. Further details not provided. | HVAC noise, telephones, office machines, conversations, and street noise. | 20-minutes sound measurements repeated 4 times at each workstation between 10:00–12:00 and 14:15–16:15. Measurements conducted on 2 separate days of similar weather conditions. | $L_{A,eq,20min}$ = 55.08 dB, SD = 4.36 dB, Range = 45.8–62.6 dB over the workstations. |
| Ayr et al (2003) [6] | Offices with between 1–6 people in Italy. Details about the kinds of rooms and office design not provided. Not clear if offices were open-plan. | HVAC noise, telephones, office machines, conversations, and external noise. | 5-minutes recordings at certain intervals during the working day at different points of the room. | $L_{A,eq,5min}$ = 44–67 dB. See paper for details about other metrics that were reported |
| Veitch et al. (2003) [14] | 9 buildings (5 public-sector in Canada, 4 private-sector in the USA and Canada), 3 with sound masking systems. Floor area: 623–3809 $m^2$. Minimum partition height: 1.5 m | Unspecified | Physical measurements for approximately 13 minutes at workstations of participants who filled in a questionnaire after relocating to a different workstation. | $L_{A,eq}$ (non-speech sounds) = 46.43 dB, SD: 3.77 dB, Range: 36.24–59.87 dB. Mean difference between A-weighted levels of low (16–500 Hz) and high frequencies (1000–8000 |

**Conclusions** (*continued*)

| Study | Room type(s) | Ambient sounds | Method | Reported values |
|---|---|---|---|---|
| | (Mean, SD: 0.2 m), Range: 0.8–2.8 m. | | Omnidirectional microphone at 1.2 m on a chair used as a portable measurement station. | Hz) = 1.96 dB, SD: 3.29 dB |
| Banbury and Berry (2005) [70] | 2 similarly sized (details not provided), large OPOs with 150 (O1) and 130 (O2) employees each in the UK. O1 had 1.5 m screens between workstation and O2 had office furniture to separate workstations. O1 occupants mostly performed secretarial-clerical tasks. O2 occupants mostly performed IT sales and online customer support. | Telephones ringing, printer, keyboard/typewriter noise and computer noise, outside noise, conversations on phones and between people. | Several 5-minutes measurements at various locations in the offices to reflect, as closely as possible, the mean ambient noise in the rooms. Each measurement was repeated an hour later, and averaged values are reported. | $L_{A,eq,5min}$ = 55 dB (O1), 60 dB (O2) |
| Kaarlela-Tuomaala et al. (2009) [41] | 4 OPOs in Finland, each 8×25×3 m (W×L×H), with 88 employees. Screens enclosed workstations on 2 or 3 sides with heights of 1.27 or 1.65 m. No acoustic treatment for floor and walls. Sound absorbing ceiling. | Voices and laughter, telephone rings, movement, doors, lift, clatter, shared office equipment, radio, HVAC, construction and traffic noise, computer sounds and vibration. | 7-hour sound measurements using dosimeters located at 1.2 m height, approximately 1–2 m to the nearest worker. Values for 15 workstations provided. | $L_{A,eq,7h}$ = 50 dB; 46–54 dB. $L_{A1}$-$L_{A99}$ = 19 dB, range ~= 13–26 dB. |
| Lenne et al. (2020) [2] | 1 OPO in France, 500 $m^2$, 83 workers. 1.4 m screens between workstations. 2 cm-thick sound absorbing ceiling tiles and ceiling-suspended absorption tiles. Walls not acoustically treated. Mostly concrete floor. | Main activity processing of customer files and limited telephonic conversations. | Sound measurements with an occupation rate of at least 75% at 3 locations in the office close to the workstations. $T$ using 2 sources and 2 receivers each | $L_{A,eq,7.30h}$ = 53.7 dB $L_{A90}$ = 41.2 dBOne-third octave band results. $T$ = 0.48–0.59 s (mean over 125–4000 Hz octave bands) |
| Park et al. (2020) [3] | 12 OPOs, 2 companies (C1, C2). 6 offices (C1) in 1 building, 6 (C2) in different buildings each. Floor area: 150–680 $m^2$; ceiling heights: 2.4–3.0 m; 10 offices with partitions with heights between 1.1–1.2 m; number of workstations 30–150. More details in paper. | Phone conversations common in most offices of C2 and included 3 call centres. | 8-hour measurements per office. Single workstation measurements in 9 offices, and measurements at 3 workstations in 3 offices | $L_{A,eq,8h}$ = 44.7–60.3 dB. $T_{20}$: 0.30–0.54 s (mean over 500 and 1000 Hz octave bands). |

## Appendix B. Supplementary data

Supplementary data to this article can be found online at https://doi.org/10.1016/j.apacoust.2021.107943.